\documentclass[longauth]{aa}  

\usepackage{graphicx}
\usepackage{txfonts}
\usepackage{hyperref}
\usepackage{euclid}
\usepackage{bm}
\usepackage{multirow}
\usepackage[dvipsnames]{xcolor}
\usepackage[inline]{enumitem}
  
\usepackage[T1]{fontenc} 
\usepackage{braket}

\newcommand{\GCs}{GC$_\textrm{s}$\xspace}
\newcommand{\GCph}{GC$_\textrm{ph}$\xspace}

\newcommand{\VIS}{m$_{\scriptscriptstyle\rm VIS}$\xspace}

\usepackage[nameinlink,noabbrev]{cleveref}
\Crefname{equation}{Eq.}{Eqs.}
\Crefname{section}{Sect.}{Sects.}
\Crefname{figure}{Fig.}{Figs.}
\crefname{equation}{Equation}{Equations}
\crefname{section}{Section}{Sections}
\crefname{figure}{Figure}{Figures}

\defcitealias{IST:paper1}{EC20}

\begin{document}

\title{\Euclid preparation. XII. Optimizing the photometric sample of the \Euclid survey for galaxy clustering and galaxy-galaxy lensing analyses}

\author{Euclid Collaboration: A.~Pocino$^{1}$\thanks{\email{pocino@ice.csic.es}}, I.~Tutusaus$^{2,3}$, F.J.~Castander$^{2,3}$, P.~Fosalba$^{2,3}$, M.~Crocce$^{1,2}$, A.~Porredon$^{2,3,4,5}$, S.~Camera$^{6,7,8}$, V.~Cardone$^{9}$, S.~Casas$^{10}$, T.~Kitching$^{11}$
% ...
}

\institute{$^{1}$ Institut de Ciencies de l'Espai (IEEC-CSIC), Campus UAB, Carrer de Can Magrans, s/n Cerdanyola del Vall\'es, 08193 Barcelona, Spain\\
$^{2}$ Institute of Space Sciences (ICE, CSIC), Campus UAB, Carrer de Can Magrans, s/n, 08193 Barcelona, Spain\\
$^{3}$ Institut d’Estudis Espacials de Catalunya (IEEC), Carrer Gran Capit\'a 2-4, 08034 Barcelona, Spain\\
$^{4}$ Department of Physics, The Ohio State University, Columbus, OH 43210, USA\\
$^{5}$ Center for Cosmology and AstroParticle Physics, The Ohio State University, 191 West Woodruff Avenue, Columbus, OH 43210, USA\\
$^{6}$ INFN-Sezione di Torino, Via P. Giuria 1, I-10125 Torino, Italy\\
$^{7}$ Dipartimento di Fisica, Universit\'a degli Studi di Torino, Via P. Giuria 1, I-10125 Torino, Italy\\
$^{8}$ INAF-Osservatorio Astrofisico di Torino, Via Osservatorio 20, I-10025 Pino Torinese (TO), Italy\\
$^{9}$ INAF-Osservatorio Astronomico di Roma, Via Frascati 33, I-00078 Monteporzio Catone, Italy\\
$^{10}$ AIM, CEA, CNRS, Universit\'{e} Paris-Saclay, Universit\'{e} de Paris, F-91191 Gif-sur-Yvette, France\\
$^{11}$ Mullard Space Science Laboratory, University College London, Holmbury St Mary, Dorking, Surrey RH5 6NT, UK\\
$^{12}$ Universit\'e Paris-Saclay, CNRS, Institut d'astrophysique spatiale, 91405, Orsay, France\\
$^{13}$ Instituto de F\'isica T\'eorica UAM-CSIC, Campus de Cantoblanco, E-28049 Madrid, Spain\\
$^{14}$ School of Physics and Astronomy, Queen Mary University of London, Mile End Road, London E1 4NS, UK\\
$^{15}$ Universit\'e St Joseph; UR EGFEM, Faculty of Sciences, Beirut, Lebanon\\
$^{16}$ Institut de Recherche en Astrophysique et Plan\'etologie (IRAP), Universit\'e de Toulouse, CNRS, UPS, CNES, 14 Av. Edouard Belin, F-31400 Toulouse, France\\
$^{17}$ INAF-Osservatorio Astronomico di Brera, Via Brera 28, I-20122 Milano, Italy\\
$^{18}$ Istituto Nazionale di Astrofisica (INAF) - Osservatorio di Astrofisica e Scienza dello Spazio (OAS), Via Gobetti 93/3, I-40127 Bologna, Italy\\
$^{19}$ IFPU, Institute for Fundamental Physics of the Universe, via Beirut 2, 34151 Trieste, Italy\\
$^{20}$ SISSA, International School for Advanced Studies, Via Bonomea 265, I-34136 Trieste TS, Italy\\
$^{21}$ INFN, Sezione di Trieste, Via Valerio 2, I-34127 Trieste TS, Italy\\
$^{22}$ INAF-Osservatorio Astronomico di Trieste, Via G. B. Tiepolo 11, I-34131 Trieste, Italy\\
$^{23}$ Universidad de la Laguna, E-38206, San Crist\'{o}bal de La Laguna, Tenerife, Spain\\
$^{24}$ Instituto de Astrof\'{i}sica de Canarias. Calle V\'{i}a L\`{a}ctea s/n, 38204, San Crist\'{o}bal de la Laguna, Tenerife, Spain\\
$^{25}$ INAF-Osservatorio di Astrofisica e Scienza dello Spazio di Bologna, Via Piero Gobetti 93/3, I-40129 Bologna, Italy\\
$^{26}$ Dipartimento di Fisica e Astronomia, Universit\'a di Bologna, Via Gobetti 93/2, I-40129 Bologna, Italy\\
$^{27}$ INFN-Sezione di Bologna, Viale Berti Pichat 6/2, I-40127 Bologna, Italy\\
$^{28}$ INAF-Osservatorio Astronomico di Padova, Via dell'Osservatorio 5, I-35122 Padova, Italy\\
$^{29}$ Universit\"ats-Sternwarte M\"unchen, Fakult\"at f\"ur Physik, Ludwig-Maximilians-Universit\"at M\"unchen, Scheinerstrasse 1, 81679 M\"unchen, Germany\\
$^{30}$ Max Planck Institute for Extraterrestrial Physics, Giessenbachstr. 1, D-85748 Garching, Germany\\
$^{31}$ Universit\'e de Paris, CNRS, Astroparticule et Cosmologie, F-75006 Paris, France\\
$^{32}$ Department of Astronomy, University of Geneva, ch. d\'Ecogia 16, CH-1290 Versoix, Switzerland\\
$^{33}$ INFN-Sezione di Roma Tre, Via della Vasca Navale 84, I-00146, Roma, Italy\\
$^{34}$ Department of Mathematics and Physics, Roma Tre University, Via della Vasca Navale 84, I-00146 Rome, Italy\\
$^{35}$ INAF-Osservatorio Astronomico di Capodimonte, Via Moiariello 16, I-80131 Napoli, Italy\\
$^{36}$ Centro de Astrof\'{\i}sica da Universidade do Porto, Rua das Estrelas, 4150-762 Porto, Portugal\\
$^{37}$ Instituto de Astrof\'isica e Ci\^encias do Espa\c{c}o, Universidade do Porto, CAUP, Rua das Estrelas, PT4150-762 Porto, Portugal\\
$^{38}$ INFN-Bologna, Via Irnerio 46, I-40126 Bologna, Italy\\
$^{39}$ Dipartimento di Fisica e Scienze della Terra, Universit\'a degli Studi di Ferrara, Via Giuseppe Saragat 1, I-44122 Ferrara, Italy\\
$^{40}$ INAF, Istituto di Radioastronomia, Via Piero Gobetti 101, I-40129 Bologna, Italy\\
$^{41}$ Universit\'{e} C\^ote d'Azur, Observatoire de la C\^ote d'Azur, CNRS, Laboratoire Lagrange, Bd de l'Observatoire, CS 34229, 06304 Nice cedex 4, France\\
$^{42}$ Instituto de Astrof\'isica e Ci\^encias do Espa\c{c}o, Faculdade de Ci\^encias, Universidade de Lisboa, Tapada da Ajuda, PT-1349-018 Lisboa, Portugal\\
$^{43}$ Observatoire de Sauverny, Ecole Polytechnique F\'ed\'erale de Lau- sanne, CH-1290 Versoix, Switzerland\\
$^{44}$ Department of Physics "E. Pancini", University Federico II, Via Cinthia 6, I-80126, Napoli, Italy\\
$^{45}$ INFN section of Naples, Via Cinthia 6, I-80126, Napoli, Italy\\
$^{46}$ INAF-Osservatorio Astrofisico di Arcetri, Largo E. Fermi 5, I-50125, Firenze, Italy\\
$^{47}$ Centre National d'Etudes Spatiales, Toulouse, France\\
$^{48}$ Institute for Astronomy, University of Edinburgh, Royal Observatory, Blackford Hill, Edinburgh EH9 3HJ, UK\\
$^{49}$ Jodrell Bank Centre for Astrophysics, School of Physics and Astronomy, University of Manchester, Oxford Road, Manchester M13 9PL, UK\\
$^{50}$ European Space Agency/ESRIN, Largo Galileo Galilei 1, 00044 Frascati, Roma, Italy\\
$^{51}$ ESAC/ESA, Camino Bajo del Castillo, s/n., Urb. Villafranca del Castillo, 28692 Villanueva de la Ca\~nada, Madrid, Spain\\
$^{52}$ Univ Lyon, Univ Claude Bernard Lyon 1, CNRS/IN2P3, IP2I Lyon, UMR 5822, F-69622, Villeurbanne, France\\
$^{53}$ Aix-Marseille Univ, CNRS, CNES, LAM, Marseille, France\\
$^{54}$ University of Lyon, UCB Lyon 1, CNRS/IN2P3, IUF, IP2I Lyon, France\\
$^{55}$ Departamento de F\'isica, Faculdade de Ci\^encias, Universidade de Lisboa, Edif\'icio C8, Campo Grande, PT1749-016 Lisboa, Portugal\\
$^{56}$ Instituto de Astrof\'isica e Ci\^encias do Espa\c{c}o, Faculdade de Ci\^encias, Universidade de Lisboa, Campo Grande, PT-1749-016 Lisboa, Portugal\\
$^{57}$ Department of Physics, Oxford University, Keble Road, Oxford OX1 3RH, UK\\
$^{58}$ INFN-Padova, Via Marzolo 8, I-35131 Padova, Italy\\
$^{59}$ Institute of Theoretical Astrophysics, University of Oslo, P.O. Box 1029 Blindern, N-0315 Oslo, Norway\\
$^{60}$ School of Physics, HH Wills Physics Laboratory, University of Bristol, Tyndall Avenue, Bristol, BS8 1TL, UK\\
$^{61}$ INAF-IASF Milano, Via Alfonso Corti 12, I-20133 Milano, Italy\\
$^{62}$ Aix-Marseille Univ, CNRS/IN2P3, CPPM, Marseille, France\\
$^{63}$ Department of Physics, P.O. Box 64, 00014 University of Helsinki, Finland\\
$^{64}$ Dipartimento di Fisica "Aldo Pontremoli", Universit\'a degli Studi di Milano, Via Celoria 16, I-20133 Milano, Italy\\
$^{65}$ INFN-Sezione di Milano, Via Celoria 16, I-20133 Milano, Italy\\
$^{66}$ Jet Propulsion Laboratory, California Institute of Technology, 4800 Oak Grove Drive, Pasadena, CA, 91109, USA\\
$^{67}$ von Hoerner \& Sulger GmbH, Schlo{\ss}Platz 8, D-68723 Schwetzingen, Germany\\
$^{68}$ Max-Planck-Institut f\"ur Astronomie, K\"onigstuhl 17, D-69117 Heidelberg, Germany\\
$^{69}$ Department of Physics and Helsinki Institute of Physics, Gustaf H\"allstr\"omin katu 2, 00014 University of Helsinki, Finland\\
$^{70}$ Universit\'e de Gen\`eve, D\'epartement de Physique Th\'eorique and Centre for Astroparticle Physics, 24 quai Ernest-Ansermet, CH-1211 Gen\`eve 4, Switzerland\\
$^{71}$ NOVA optical infrared instrumentation group at ASTRON, Oude Hoogeveensedijk 4, 7991PD, Dwingeloo, The Netherlands\\
$^{72}$ Argelander-Institut f\"ur Astronomie, Universit\"at Bonn, Auf dem H\"ugel 71, 53121 Bonn, Germany\\
$^{73}$ Institute for Computational Cosmology, Department of Physics, Durham University, South Road, Durham, DH1 3LE, UK\\
$^{74}$ Universit\'{e} de Paris, F-75013, Paris, France, LERMA, Observatoire de Paris, PSL Research University, CNRS, Sorbonne Universit\'{e}, F-75014 Paris, France\\
$^{75}$ California institute of Technology, 1200 E California Blvd, Pasadena, CA 91125, USA\\
$^{76}$ INAF-IASF Bologna, Via Piero Gobetti 101, I-40129 Bologna, Italy\\
$^{77}$ Institut de F\'{i}sica d’Altes Energies (IFAE), The Barcelona Institute of Science and Technology, Campus UAB, 08193 Bellaterra (Barcelona), Spain\\
$^{78}$ Institute of Cosmology and Gravitation, University of Portsmouth, Portsmouth PO1 3FX, UK\\
$^{79}$ European Space Agency/ESTEC, Keplerlaan 1, 2201 AZ Noordwijk, The Netherlands\\
$^{80}$ ICC\&CEA, Department of Physics, Durham University, South Road, DH1 3LE, UK\\
$^{81}$ Department of Physics and Astronomy, University of Aarhus, Ny Munkegade 120, DK–8000 Aarhus C, Denmark\\
$^{82}$ Perimeter Institute for Theoretical Physics, Waterloo, Ontario N2L 2Y5, Canada\\
$^{83}$ Department of Physics and Astronomy, University of Waterloo, Waterloo, Ontario N2L 3G1, Canada\\
$^{84}$ Centre for Astrophysics, University of Waterloo, Waterloo, Ontario N2L 3G1, Canada\\
$^{85}$ Space Science Data Center, Italian Space Agency, via del Politecnico snc, 00133 Roma, Italy\\
$^{86}$ Institute of Space Science, Bucharest, Ro-077125, Romania\\
$^{87}$ Institute for Computational Science, University of Zurich, Winterthurerstrasse 190, 8057 Zurich, Switzerland\\
$^{88}$ Dipartimento di Fisica e Astronomia “G.Galilei", Universit\'a di Padova, Via Marzolo 8, I-35131 Padova, Italy\\
$^{89}$ Departamento de F\'isica, FCFM, Universidad de Chile, Blanco Encalada 2008, Santiago, Chile\\
$^{90}$ INFN-Sezione di Roma, Piazzale Aldo Moro, 2 - c/o Dipartimento di Fisica, Edificio G. Marconi, I-00185 Roma, Italy\\
$^{91}$ Institut d'Astrophysique de Paris, 98bis Boulevard Arago, F-75014, Paris, France\\
$^{92}$ Universidad Polit\'ecnica de Cartagena, Departamento de Electr\'onica y Tecnolog\'ia de Computadoras, 30202 Cartagena, Spain\\
$^{93}$ Kapteyn Astronomical Institute, University of Groningen, PO Box 800, 9700 AV Groningen, The Netherlands\\
$^{94}$ Department of Physics, P.O.Box 35 (YFL), 40014 University of Jyv\"askyl\"a, Finland\\
$^{95}$ Infrared Processing and Analysis Center, California Institute of Technology, Pasadena, CA 91125, USA\\
$^{96}$ Department of Physics and Astronomy, University College London, Gower Street, London WC1E 6BT, UK\\
}

\date{}

\authorrunning{Euclid Collaboration: A. Pocino et al.}
\titlerunning{Optimizing the photometric sample of the \Euclid survey for \GCph and GGL analyses}

   \abstract{Photometric redshifts (photo-$z$s) are one of the main ingredients in the analysis of cosmological probes. Their accuracy particularly affects the results of the analyses of galaxy clustering with photometrically-selected galaxies (\GCph) and weak lensing. In the next decade, space missions like \Euclid will collect precise and accurate photometric measurements for millions of galaxies. These data should be complemented with upcoming ground-based observations to derive precise and accurate photo-$z$s. In this article we explore how the tomographic redshift binning and depth of ground-based observations will affect the cosmological constraints expected from the \Euclid mission. We focus on \GCph and extend the study to include galaxy-galaxy lensing (GGL). We add a layer of complexity to the analysis by simulating several realistic photo-$z$ distributions based on the Euclid Consortium Flagship simulation and using a machine learning photo-$z$ algorithm. We then use the Fisher matrix formalism presented in \citet{IST:paper1} together with these galaxy samples to study the cosmological constraining power as a function of redshift binning, survey depth, and photo-$z$ accuracy. We find that bins with equal width in redshift provide a higher Figure of Merit (FoM) than equipopulated bins and that increasing the number of redshift bins from 10 to 13 improves the FoM by $35\%$ and $15\%$ for \GCph and its combination with GGL, respectively. For \GCph, an increase of the survey depth provides a higher FoM. However, when we include faint galaxies beyond the limit of the spectroscopic training data, the resulting FoM decreases because of the spurious photo-$z$s. When combining \GCph and GGL, the number density of the sample, which is set by the survey depth, is the main factor driving the variations in the FoM. Adding galaxies at faint magnitudes and high redshift increases the FoM even when they are beyond the spectroscopic limit, since the number density increase compensates the photo-$z$ degradation in this case. We conclude that there is more information that can be extracted beyond the nominal 10 tomographic redshift bins of \Euclid and that we should be cautious when adding faint galaxies into our sample, since they can degrade the cosmological constraints.}

   \keywords{}

   \maketitle
%

\section{Introduction}

The goal of Stage-IV dark energy surveys \citep{Albrecht_2006}, such as \Euclid\footnote{\url{https://www.euclid-ec.org}} \citep{Laureijs_2011} and the Vera C. Rubin Observatory Legacy Survey of Space and Time,\footnote{\url{https://www.lsst.org}} (Rubin-LSST; \citealt{LSST_2009}) is to measure both the expansion rate of the Universe and the growth of structures up to redshift $z\sim 2$ and beyond. These surveys will allow us to constrain a large variety of cosmological models using cosmological probes like weak gravitational lensing (WL) and galaxy clustering. Stage-IV surveys can be classified into spectroscopic and photometric surveys, depending on whether the redshift of the observed objects is estimated with spectroscopy or using photometric techniques. The latter can provide measurements for many more objects than the former but at the expense of a degraded precision on the redshift estimates, given that photometric surveys observe through multi-band filters instead of observing the full spectral energy distribution that requires more observational time. Because of this, galaxy clustering analyses are usually performed with data coming from spectroscopic surveys, while the data obtained from photometric surveys are generally used for WL analyses. However, given the current (and future) precision of our measurements, the signal we can extract from galaxy clustering analyses using photometric surveys is far from being negligible \citep[see e.g.][]{DESY1-GCWL,vanUitert:2017ieu,IST:paper1,Tutusaus2020}. Therefore, upcoming surveys can increase their constraining power if they optimize their photometric samples to include galaxy clustering studies in addition to WL analyses.
The main aim of this work is to perform such an optimization study for the \Euclid photometric sample.

The \Euclid satellite will observe over a billion galaxies through an optical and three near-infrared broad bands. 
Given the specifications of the satellite, the combination of \Euclid and ground-based surveys can enrich the science exploitation of both. On one hand, the WL analysis of \Euclid data requires an accurate knowledge of the redshift distributions of the samples used for the analysis. \Euclid photometric data alone cannot reach the necessary photometric redshift (photo-$z$) performance and additional ground-based data are required. 
On the other hand, 
\Euclid will provide additional information to ground-based surveys such as very precise shape measurements -- thanks to the high spatial resolution achieved being in space and avoiding atmospheric distortions  -- and near-infrared spectroscopy. \Euclid's data will help ground-based surveys improve their deblending of faint objects and improve their photo-$z$ estimates, which will definitely boost their scientific outcome. 
Surveys where these synergies can be established include
 the Panoramic Survey Telescope and Rapid Response System\footnote{\url{https://panstarrs.stsci.edu}} (PanSTARRS; \citealt{Chambers_2016}), the Canada-France Imaging Survey\footnote{\url{http://www.cfht.hawaii.edu/Science/CFIS/}} (CFIS; \citealt{CFIS}), the Hyper Suprime-Cam Subaru Strategic Program\footnote{\url{https://hsc.mtk.nao.ac.jp/ssp/}} (HSC-SSP; \citealt{HSC}),
 the Javalambre-\Euclid Deep Imaging Survey (JEDIS), the Dark Energy Survey\footnote{\url{https://www.darkenergysurvey.org}} (DES; \citealt{DES_2005}), or 
 Rubin-LSST \citep{Ivezic_2019}.
The latter is a Stage IV experiment with a strong complementarity with \Euclid
since it greatly overlaps in area, covers two \Euclid Deep Fields and reaches a faint photometric depth that will lead to better photo-$z$ estimation \citep{Rhodes_2017, Capak_2018}. In this article we consider the addition of ground-based optical photometry to \Euclid in order to assess the optimal photometric sample for galaxy clustering and galaxy-galaxy lensing (GGL) analyses.

The optimization of the sample of photometrically-selected galaxies for galaxy clustering analyses has been already studied in the literature. In \citet{Tanoglidis_2019}, the authors focus their analysis on galaxy clustering for the first three years of DES data. Also for DES but including galaxy-galaxy lensing, \citet{Porredon_2021} studies lens galaxy sample selections based on magnitude cuts as a function of photo-$z$, balancing density and photo-$z$ accuracy to optimise cosmological constrains in the wCDM space.
Another example is the recent analysis of \citet{Eifler_2020_b} on the Nancy Grace Roman Space Telescope \citep{Spergel_2015} High Latitude Survey (HLS), where the authors simulate and explore multi-cosmological probes strategies on dark energy and modified gravity to study observational systematics, such as photo-$z$. These studies show the importance of optimizing the galaxy sample for galaxy clustering analysis. We aim to perform a similar optimization for the \Euclid mission. 
Note that there have also been several studies optimizing the spectroscopic sample for galaxy clustering analysis with \Euclid \citep{Samushia_2011, Wang_2010}.

We want to optimize the \Euclid sample of galaxies detected with photometric techniques by performing realistic forecasts of its cosmological performance and observing the improvement on the cosmological constraining power of different galaxy samples. When performing galaxy clustering analyses with a photometric sample there are several effects that need to be taken into account such as galaxy bias, photo-$z$ uncertainties or shot noise, among other effects. 
Here, we try to follow the procedures one would perform in a real data analysis when  selecting the samples for the analysis. For that purpose, we use the \Euclid Flagship simulation (Euclid Collaboration, in prep; \citealt{Potter_2017}). For a given expected limit of the photometric depth, we select the galaxies included within that magnitude limit and use a machine learning photo-$z$ method to study the optimal way to split the catalogue into subsamples for the analysis. We generate realistic redshift distributions, $n(z)$, for the chosen subsamples and estimate their galaxy bias, $b(z)$.
We study the constraining power of these samples when we modify the number and width of the tomographic bins, and when we reduce the sample size by performing a series of cuts in magnitude.

The article is organized as follows. We present \Euclid and ground-based surveys in \Cref{sec:Euclid_survey} and \Cref{sec:Ground_based_surveys}. In \Cref{sec:Realistic_photometric_redshift} we introduce the Flagship simulation and describe how we create photometric samples with different selection criteria. We define the set of galaxy samples that will be used throughout the article, and explain how we estimate the photometric redshifts. In \Cref{sec:building-forecasts} we detail the forecast formalism and we describe the cosmological model in \Cref{sec:cosmo}. In \Cref{sec:results} we present the results of the optimization when changing the number and type of tomographic bins, and we study the dependency of the cosmological constraints on photo-$z$ quality and sample size. Finally, we present our conclusions in \Cref{sec:conclusion}.


\section{The \Euclid survey}\label{sec:Euclid_survey}

\Euclid is an European Space Agency (ESA) M-class space mission due for launch in 2022. In the wide survey, it will cover over $15\,000$ deg$^2$ of the extra-galactic sky with the main aim of measuring the geometry of the Universe and the growth of structures up to redshift $z\sim2$ and beyond. \Euclid will have two instruments on-board: a near-infrared spectro-photometer \citep{NISP_paper}, and an imager at visible wavelengths \citep{VIS_paper}. 
The imager of \Euclid, called VIS, will observe galaxies through an optical broad band, \VIS, covering a wavelength range between $540$ and $900$ nm, with a magnitude depth of 24.5 at $10\sigma$ for extended sources. The spectro-photometric instrument, called NISP, has three near-infrared bands, $YJH$, covering a wavelength range between $920$ and $2000$ nm \citep{Racca_2016, Racca_2018}. The nominal survey exposure is expected to reach a magnitude depth of 24 at $5\sigma$ for point sources. If we convert this depth  to $10\sigma$ level detections for extended sources we obtain a magnitude depth of about 23, which is the value we consider in \Cref{tab:Coadded_depths_10sigma}. 
The deep survey will cover $40$ deg$^{2}$ divided in three different fields: the Euclid Deep Field North and the Euclid Deep Field Fornax of 10 deg$^{2}$ each, and the Euclid Deep Field South of $20$ deg$^{2}$ (Euclid Collaboration, in prep.). In these fields the magnitude depth will be two magnitudes deeper than in the wide survey.
With its two instruments, \Euclid will perform both a spectroscopic and a photometric galaxy survey that will allow us to determine cosmological parameters using its three main cosmological probes: galaxy clustering with the spectroscopic sample (\GCs), galaxy clustering with the photometric sample (\GCph), and WL. 
We will study how the selection of the galaxy sample that enters into the analysis can be optimised to provide the tightest cosmological constraints focusing on the \GCph analysis and its cross-correlation with WL -- also called GGL.


\section{Ground-based surveys}
\label{sec:Ground_based_surveys}

The single broad band VIS of \Euclid cannot sample the spectral energy distribution in the optical range. \Euclid will require complementary observations in the optical from ground-based surveys to provide the photometry to estimate accurate photometric redshifts and achieve the scientific goals of \Euclid. Several ground-based surveys will be needed to cover all the observed area of \Euclid, as \Euclid covers both celestial hemispheres and those cannot be reached from a single observatory on Earth. The ground-based complementary data will not cover uniformly the \Euclid footprint. It is very likely that there will be at least three distinct areas in terms of photometric data available. The Southern hemisphere is expected to be covered with Rubin-LSST data, while the Northern hemisphere will be covered with a combination of surveys such as CFIS, PanSTARRS, JEDIS and HSC-SPP.  
In addition, some area North of the equator may also be covered by Rubin-LSST at a shallower depth than in the Southern hemisphere.
In this work we include simulated ground-based photometry 
that try to encompass the range of possible ground-based depths that the \Euclid analysis will have from the deepest Rubin-LSST data to the shallower data from other surveys.


Rubin-LSST is expected to start operations in 2022 and over 10 years it will observe over $20\,000$ deg$^2$ in the Southern hemisphere with 6 optical bands, $ugrizy$, covering a wavelength range from $320$ to $1050$ nm. The idealized final magnitude depth for coadded images for $5\sigma$ point sources  are 26.1, 27.4, 27.5, 26.8, 26.1, 24.9, for $ugrizy$, respectively, based on the Rubin-LSST design specifications~\citep{Ivezic_2019}.
Among other scientific themes, Rubin-LSST has been designed to study dark matter and dark energy using WL, \GCph, 
and supernovae as cosmological probes.
The Rubin-LSST survey will provide the best photometry for \Euclid-detected galaxies at the time that \Euclid data become available.  

Another suitable ground-based candidate to cover the optical and near-infrared range in the Southern sky is the DES photometric survey. DES completed observations in 2019 after a 6-years program. It covered $5000$ deg$^{2}$ around the Southern Galactic cap through 5 broad band filters, $grizy$, with wavelength ranging from $400$ to $1065$ nm, and redshift up to $1.4$ \citep{DES_2016}. The median coadded magnitude limit depths for 10$\sigma$ and $\ang{;;2}$ 
diameter aperture are 
24.3, 24.0, 23.3, 22.6, for $griz$, respectively. These depths correspond to the published values of the first three years of observations \citep{Sevilla-Noarbe_2020}.


\section{Generating realistic photometric galaxy samples}\label{sec:Realistic_photometric_redshift}

The cosmological constraining power of \Euclid will depend on the external data available as it will dictate the photo-$z$ performance of the samples to be studied. In order to study the impact of the available photometry, we create six samples selected with different photometric depths. For each sample, we compute the photo-$z$ estimates using machine learning techniques taking into account the expected spectroscopic redshift distribution of the training sample. We use these photo-$z$ estimates to split each sample into tomographic bins for which we can compute their photo-$z$ distributions and galaxy bias from the simulation. These $n(z)$ and $b(z)$ are then used to forecast the cosmological performance.
In this section we provide a detailed description of how we obtain the realistic photo-$z$ estimates of the \Euclid galaxies that are later used in the forecast. We first present the cosmological simulation used to extract the photometry and the galaxy distributions. We then explain how we generate realisations of the photometry for the simulated galaxies taking into account the expected depth of the \Euclid and ground-based data. We finally present the method used to estimate the photo-$z$.

\subsection{The Flagship simulation}\label{sec:Flagship}

We consider the Flagship galaxy mock catalogue of the Euclid Consortium (Euclid Collaboration, in preparation) 
to create the different samples. The catalogue uses the Flagship N-body dark matter simulation \citep{Potter_2017}. Dark matter halos are identified using ROCKSTAR \citep{Behroozi_2013} and are retained down to a mass of $2.4\times10^{10}$ h$^{-1}\,\textup{M}_\odot$, which corresponds to ten particles. Galaxies are assigned to dark matter halos using Halo Abundance Matching (HAM) and Halo Occupation Distribution (HOD) techniques. The cosmological model assumed in the simulation is a flat $\Lambda$CDM model with fiducial values
$\Omega_{\rm m} = 0.319$, $\Omega_{\rm b}=0.049$, $\Omega_{\Lambda}=0.681$, $\sigma_{8}=0.83$, $n_{\rm s}=0.96$, $h=0.67$. The N-body simulation ran in a $3.78$ h$^{-1}$ Gpc box with particle mass $m_{\rm p}=2.398\times10^{9}$ h$^{-1}\,\textup{M}_\odot$.
The galaxy mock generated has been calibrated using local observational constraints, such as the luminosity function from \citet{Blanton_2003} and \citet{Blanton_2005} for the faintest galaxies, the galaxy clustering measurements as a function of luminosity and colour from \citet{Zehavi_2011}, and the colour-magnitude diagram as observed in the New York University Value Added Galaxy Catalog \citep{Blanton_2005_b}. The catalogue contains about $3.4$ billion galaxies over $5000$ deg$^{2}$ and extends up to redshift $z = 2.3$. 

For this study we select an area of $402$ deg$^{2}$, which corresponds to galaxies within the range of right ascension $15^{\circ}<\alpha<75^{\circ}$ and declination $62^{\circ}<\delta<90^{\circ}$. All the photometric galaxy distributions obtained in this patch are extrapolated to the $15\,000$ deg$^{2}$ of sky that \Euclid is expected to observe. Note that the selected area is large enough to minimize the impact of sample variance, but small enough to allow for the production of several galaxy samples in a reasonable amount of time. After the photometric uncertainty is added to the photometry of each galaxy, we perform a magnitude cut in \VIS $<25$ that leads to a number density of about $41.5$ galaxies per arcmin$^{2}$.

\subsection{Photometric depth}\label{sec:Photometry_degradation}

Each galaxy observation will lead to a measured value of its magnitude and its associated error. The magnitude depth is usually given as the magnitude at which the median relative error has a particular value. In galaxy surveys it is customary to express the depth at a signal-to-noise of 10 for extended objects, that is, when the value of the noise is one tenth of its signal. As explained in detail below, we generate realisations of the photometric errors for a given survey taking into account its magnitude depth and scaling the values of the errors at other magnitudes assuming background limited observations, that is, that the background signal dominates the contribution to the error.


We simulate four different photometric survey depths. \Cref{tab:Coadded_depths_10sigma} shows their magnitude limits. The first column corresponds to a combination of \Euclid and ground-based photometric depth expected to be achieved in the Southern hemisphere. We label this case as optimistic and it is the deepest case we will study. The magnitude limits for the optical bands are for extended sources at $10\sigma$, similar to those expected from Rubin-LSST \citep{LSST_Science_Collaboration_2009}.
The values for \Euclid correspond to a $10\sigma$ detection level for extended sources.
In addition to the magnitude limits expected in the South, we also want to investigate how the cosmological constraints degrade as the depth is reduced. We investigate three other cases. First, a case were the depth in optical bands are reduced by a factor of two in signal-to-noise ratio. The second column shows the magnitudes limits for this case where the optical bands are reduced by 0.75 magnitude.
This column represents a possible case where the Rubin-LSST data have a reduced depth in areas outside its main footprint.
We also study a case were the limiting fluxes of \Euclid are brightened by 0.75 magnitudes, shown in the third column. Lastly, we explore a case where the ground-based data is degraded by a factor of five in signal-to-noise but the \Euclid space data remains at their nominal depth values. This broadly represents the depth that can be achieved from other ground-based data in the Northern hemisphere.

For each survey case, we generate a galaxy catalogue drawn from the Flagship simulation. We assign observed magnitudes and errors with the following procedure. First, we compute the expected error for each galaxy, taking into account its magnitude in the Flagship catalogue and the magnitude limit of the survey as given in~\Cref{tab:Coadded_depths_10sigma}. We assume that the observations are sky limited (the noise is dominated by the shot noise of the sky), and therefore we scale the ratio of the signal to noise between two galaxies $i$ and $j$ as the ratio of their fluxes
\begin{equation}
    \left(\frac{S}{N}\right)_{i} = \left(\frac{S}{N}\right)_{j}\frac{f_{i}}{f_{j}}\,,
\end{equation}
where $f_{i}$ is the observed flux of galaxy $i$ detected at signal-to-noise ratio $\left(S/N\right)_{i}$. The magnitude (flux) limits in~\Cref{tab:Coadded_depths_10sigma} give us the fluxes corresponding to a signal-to-noise ratio of 10, $f_{10\sigma}$, and therefore we can compute the expected signal-to-noise at which a galaxy of a given magnitude is detected as
\begin{equation}
    \left(\frac{S}{N}\right)_{i} =
    10\frac{f_{i}}{f_{10\sigma}}\,.
\end{equation}
Using the definition of signal-to-noise, $ (S/N)_{i} = f_{i}/\Delta f_{i}$, we can compute the expected flux error for each galaxy as
\begin{equation}\label{eq:Flux_error}
    \Delta f_{i}=\frac{f_{10\sigma}}{10}\,.
\end{equation}
The fluxes in the Flagship catalogue correspond to the real fluxes of each galaxy. Whenever we observe these galaxies in a given survey, we detect a realization of the real flux. For our study, we generate realisations of the observed fluxes $f_{i}^{*}$ for each survey as
\begin{equation}
f_{i}^{*}=f_{i} + N\left(\mu = 0,\,\sigma=f_{10\sigma}/10\right)\,,
\end{equation}
where $N$ is a random number from a normal distribution. We then assign errors to the resulting fluxes according to~\Cref{eq:Flux_error}. Finally, the new fluxes and their assigned errors are converted into magnitudes and their respective magnitude errors. 

\begin{table*}
	\centering
	\caption{Limiting coadded depth magnitudes for extended sources at $10\sigma$ used in each sample. 
	}
	\label{tab:Coadded_depths_10sigma}
	\begin{tabular}{lccccc}
		\hline
		\hline
		\noalign{\smallskip}
		& & & Ground based & All & Ground based\\
		\multicolumn{2}{c}{Band} & Optimistic &  degraded $ - 0.75 $ & degraded $ - 0.75 $ & degraded $-1.75$\\
        \noalign{\smallskip}
		\hline
		\noalign{\smallskip}
		& $u$ & 25.55 & 24.8 & 24.8 & 23.8\\
		& $g$ & 26.75 & 26.0 & 26.0 & 25.0\\
		Ground & r & 26.95 & 26.2 & 26.2 & 25.2\\
		based & i & 26.25 & 25.5 & 25.5 & 24.5\\
		& $z$ & 25.45 & 24.7 & 24.7 & 23.7\\
		& $y$ & 24.15 & 23.4 & 23.4 & 22.4\\
		\noalign{\smallskip} \hline \noalign{\smallskip}
		\multicolumn{1}{c}{\multirow{4}{*}{\Euclid}} & \VIS & 24.6 & 24.6 & 23.85 & 24.6\\
		& $Y$ & 23 & 23 & 22.25 & 23\\
		& $J$ & 23 & 23 & 22.25 & 23\\
		& $H$ & 23 & 23 & 22.25 & 23\\
		\noalign{\smallskip}
		\hline
		\hline
	\end{tabular}
\end{table*}

\subsection{Samples}\label{sec:Samples}

\begin{figure}
	\includegraphics[width=\columnwidth]{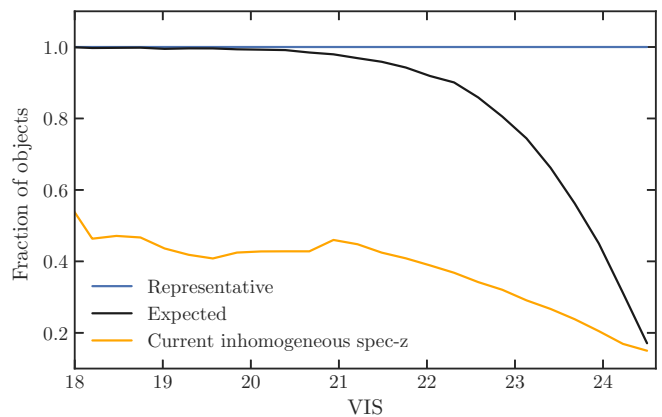}
    \caption{Fraction of simulated objects with successful spectroscopic redshift as a function of \VIS. The lines represent the completeness fraction 
    of the spectroscopic training samples. The blue line corresponds to the fraction of objects for a random training subsample that is fully representative of the sample under study.
    In black we show an expectation of the spectroscopic completeness for future ground-based surveys such as Rubin-LSST in \VIS (see \citealt{Newman_2015}). In orange we present the completeness of a training sample with an $n(z)$ similar to the currently available spectroscopic data (see text). 
    Note that the number of objects included in each training set is not represented by the normalisation of the different curves in this figure (see~\Cref{fig:samples_training_redshift_distributions} for the redshift distributions). 
    Moreover, although our photometric samples go up to \VIS $=25$, we cut the spectroscopic training samples at \VIS $<24.5$ because realistic redshifts have not been reliably determined beyond that magnitude limit yet.}
    \label{fig:completeness}
\end{figure}

We estimate the expected cosmological constraints using the galaxy clustering analysis of tomographic bins defined with photo-$z$ (see \Cref{sec:building-forecasts}). The magnitude limit of a given sample will give us the galaxies that form the overall sample, while the photo-$z$ algorithm will split that sample into tomographic bins and will provide an estimate of the redshift distributions within these tomographic bins. We can better understand the uncertainties in the method using simulations where we know the true redshift distributions. So far, we have defined four different samples based on the available photometry representing the four cases defined in~\Cref{tab:Coadded_depths_10sigma}. The photo-$z$ performance depends on the photometric depth and the spectroscopic data available to train the method. Now, we will generate study cases depending on the spectroscopic data available to train the photo-$z$ technique we use. We will use three different spectroscopic samples with different completeness profiles as a function of magnitude. First, we consider an idealised case where the spectroscopic training sample is a random subsample of the whole sample and thus it is fully representative (blue line in \Cref{fig:completeness}). We consider a second case where the spectroscopic sample completeness as a function of magnitude follows the  expectations from spectrographs on 8-m class telescopes \citep{Newman_2015}. This case is shown in black in \Cref{fig:completeness}. 
This is intended to mimic the spectroscopic incompleteness as a function of magnitude of surveys like zCOSMOS \citep{Lilly_2007}, VVDS \citep{LeFevre_2013} and DEEP2 \citep{Newman_2013} at least in its shape, although maybe optimistic in its normalisation. Finally, we consider a last case where the spectroscopic completeness is similar to the current available spectroscopic surveys, as those listed in \citet{Gschwend_2018}. We compute how the completeness in spectroscopic data as a function of redshift translates into completeness in \VIS (orange line in \Cref{fig:completeness}). These cases are explained in more detail later in this section.
It is worth mentioning that we only consider galaxies and not stars in the samples under study. With the high spatial resolution of Euclid, the contamination in the sample due to stars is expected to be minimal. We have also assumed that the effects of Galactic extinction are corrected in the data reduction pipelines and therefore ignore Galactic extinction. These factors can be include in the future to add another layer of realism to the analysis.

\begin{table}
	\centering
	\caption{Cases under study. The photometric limit value corresponds to the column number of \Cref{tab:Coadded_depths_10sigma} whose magnitude limit depths are used to define each photometric sample. The spectroscopic training sample used to determine the photo-$z$ can be a representative subsample, a sample with a completeness drop in \VIS or a sample with an inhomogeneous spectroscopic redshift distribution as shown in \Cref{fig:completeness}.}
	\label{tab:cases}
	\begin{tabular}{lcl}
		\hline
		\hline
		\noalign{\smallskip}
		\multicolumn{1}{l}{\multirow{2}{*}{Sample name}} & Photometric & Spectroscopic\\
		& limit & training\\
        \noalign{\smallskip}
		\hline
		\noalign{\smallskip}
		Case 1: Optimistic & 1 & Subsample \\
		Case 2: Fiducial & 1 & Compl. drop \\
		Case 3: Mid-depth & 2 & Compl. drop \\
		Case 4: Mid-depth \Euclid & 3 & Compl. drop \\
		Case 5: Shallow depth & 4 & Compl. drop \\
		Case 6: Inhomogeneous spec & 4 & Inho. spec-z\\
		\noalign{\smallskip}
		\hline
		\hline
	\end{tabular}
\end{table}

We combine the four cases of photometric limits with the three cases of different spectroscopic data available to train the photo-$z$ techniques to generate six galaxy samples for our study. With these six samples we try to encompass a wide range of scenarios to try to understand how the cosmological constraints vary depending on the sample available. We detail these six cases in the following subsections. \Cref{tab:cases} summarises all the cases we consider.
All our samples have galaxies down to a magnitude limit of \VIS $= 25$. Note that for our shallower survey (column four in~\Cref{tab:Coadded_depths_10sigma}), galaxies near this \VIS selection limit will have larger errors.  It is also important to mention that in all cases we assume the magnitude limit in each band to be isotropic -- homogeneous on the sky. This will definitely not be the case for \Euclid, since ground-based data will consist on a compilation of different surveys pointing at different regions of the sky, with different depths and systematic uncertainties. For instance, Rubin-LSST focuses on the Southern hemisphere, while \Euclid will also observe the Northern one. A more detailed analysis taking into account the depth anisotropy of the ground-based data is left for future work. A possible approach would be to generate several sets of ground-based photometry according to the specific limitations of each ground-based instrument and region of the sky covered, in order to reproduce the expected anisotropy of the photometry. Then we would mix the different sets of ground-based photometry and add them to the \Euclid photometry in order to determine the photometric redshifts and redo the optimization analysis as performed in this article.

\subsubsection{Case 1: Optimistic}

This case uses the deepest magnitude limit and a highly idealised spectroscopic training sample. The sample has magnitudes and errors generated as described in~\Cref{sec:Photometry_degradation} with the \Euclid and ground-based photometric depth limits shown in the first column of~\Cref{tab:Coadded_depths_10sigma}. 
The photo-$z$ are estimated using a training set that is a complete and representative subsample in both redshift and magnitude of the whole sample.

\subsubsection{Case 2: Fiducial}

We take this case to be our fiducial sample. We use the deepest photometry as in the optimistic case~1 but the photo-$z$ estimation now makes use of a training sample that has a completeness drop at faint magnitudes that resembles the incompleteness of spectroscopic surveys carried out with spectrographs in 8m-class telescopes.
We show the completeness drop in the spectroscopic training sample in \Cref{fig:completeness} (black line). 
While the completeness as a function of magnitude intends to be realistic of current spectroscopic capabilities, we make the simplifying assumption that this incompleteness does not depend on any galaxy property except its magnitude and therefore we randomly subsample the whole distribution only taking into account the probability of being selected based on the galaxy magnitude.

\subsubsection{Case 3: Ground-based mid-depth photometry}

We define another sample trained with the same spectroscopic training sample completeness as in the fiducial case but with shallower ground-based magnitude limits in the photometry. The ground-based magnitude limit is a factor of two shallower in signal-to-noise ratio than in cases 1--2. This corresponds to the second column in \Cref{tab:Coadded_depths_10sigma}. This case is intended to represent areas on the sky between the celestial equator and low Northern declinations where Rubin-LSST data at shallower depth may be available.

\subsubsection{Case 4: \Euclid mid-depth photometry}

To explore the possibilities of available photometry, especially the importance of deep near-infrared photometry, we define a case in which both the \Euclid and ground-based photometric depth is reduced by 0.75 magnitudes (third column in \Cref{tab:Coadded_depths_10sigma}). The spectroscopic training sample completeness is the same as in cases 2 and 3.

\subsubsection{Case 5: Ground-based shallow depth photometry}

The complementary ground-based photometry expected to be available in the Northern hemisphere is shallower than the magnitude limits used in our previous cases.
We define a sample to roughly represent and cover this option by considering a ground-based flux limit 1.75 magnitudes brighter compared to our optimistic case (fourth column in \Cref{tab:Coadded_depths_10sigma}). To compute the photo-$z$ we use a spectroscopic training set with the same completeness in \VIS as in cases 2, 3, and 4.

\subsubsection{Case 6: Inhomogeneous spectroscopic sample}

In this last sample, we want to study the case in which the spectroscopic training sample is very heterogeneous and composed of the combination of many surveys targeting galaxies with different selection criteria and with different spectroscopic facilities. We choose a spectroscopic training set that tries to model the $n(z)$ of current available spectroscopic data coming from surveys as those listed in \citet{Gschwend_2018}.
Given that some of these surveys have different colour selection cuts and magnitude limit depths, the combined redshift distribution is not homogeneous presenting peaks and troughs, which cause strong biases in the photo-$z$ estimation due to over and under-represented galaxies at different redshift ranges (see e.g. \citealt{Zhou_2020}).
We want to remark that we only try to reproduce the $n(z)$ of the overall spectroscopic sample. We do not try to gather this spectroscopic sample applying the same selection criteria of the different surveys used.
We consider that this is not necessary for our purposes as we are only interested in the overall trend induced by using an inhomogeneous spectroscopic training sample. We create the spectroscopic training sample by randomly selecting galaxies based on their redshift to reproduce the overall targeted redshift distribution. 
Given that the Flagship simulation area we are using (see \Cref{sec:Flagship}) is smaller than the surveys sampling the nearby Universe, our simulated spectroscopic training does not exactly reproduced our overall redshift distribution at low redshifts.
The resulting completeness as a function of the \VIS of this spectroscopic redshift sample can be seen in \Cref{fig:completeness} (orange line). The modeled $n(z)$ is shown in \Cref{fig:samples_training_redshift_distributions} (orange line).
With this case, which intends to represent the currently available data, we can draw a lower bound on the photo-$z$ accuracy that can be expected for \Euclid.
In this case, we use the same photometric magnitude limits as in case 5.

The realism of our training samples is limited in the sense that we only try to reproduce the completeness in \VIS or the shape of the $n(z)$ distribution. We do not take into account any dependence of the training samples on other characteristics such as galaxy type or the presence of emission lines, which would have an impact on the determination of the photo-$z$.

\subsection{Photometric redshifts}\label{sec:Photometric_redshift}

\begin{figure}
	\includegraphics[width=\columnwidth]{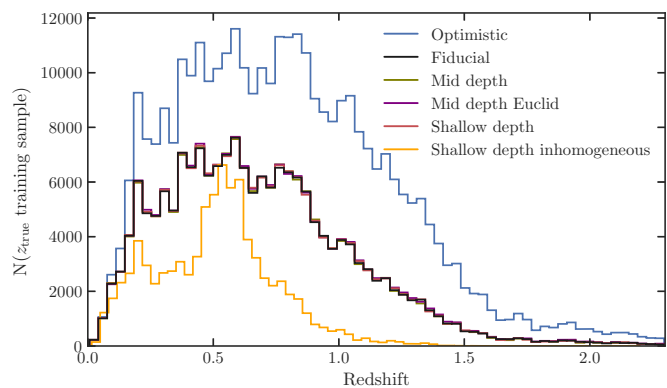}
    \caption{True redshift distributions of the training samples used to run DNF in all 6 cases. The training samples include magnitudes brighter than \VIS $=24.5$. The true redshift comes from the Flagship simulation. 
    The four training samples with almost identical true redshift distributions have the same completeness drop in \VIS and only differ in the photometric quality.}
    \label{fig:samples_training_redshift_distributions}
\end{figure}
\begin{figure}
	\includegraphics[width=\columnwidth]{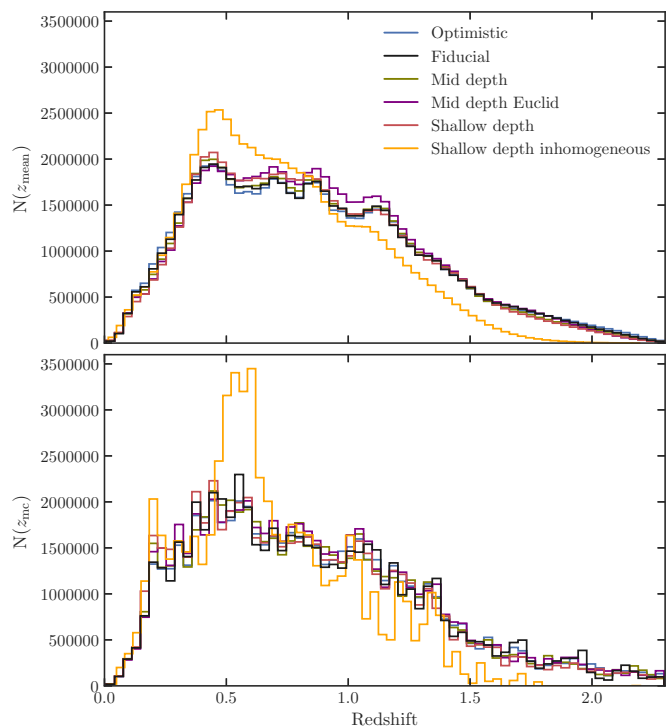}
    \caption{
    \textit{Top:} $z_{\rm mean}$ photometric redshift distributions obtained with DNF for the 6 photometric samples up to \VIS $=25$. The $z_{\rm mean}$ photo-$z$ estimate returned by DNF is the value resulting from the mean of the nearest neighbours redshifts. 
    \textit{Lower:} Photometric redshift distributions obtained with DNF for the $z_{\rm mc}$ statistic, which for each galaxy is a one-point sampling of the redshift probability distribution estimated from the nearest neighbour (see text for details).}
    \label{fig:samples_redshift_distributions}
\end{figure}

\begin{figure*}
	\includegraphics[width=\textwidth]{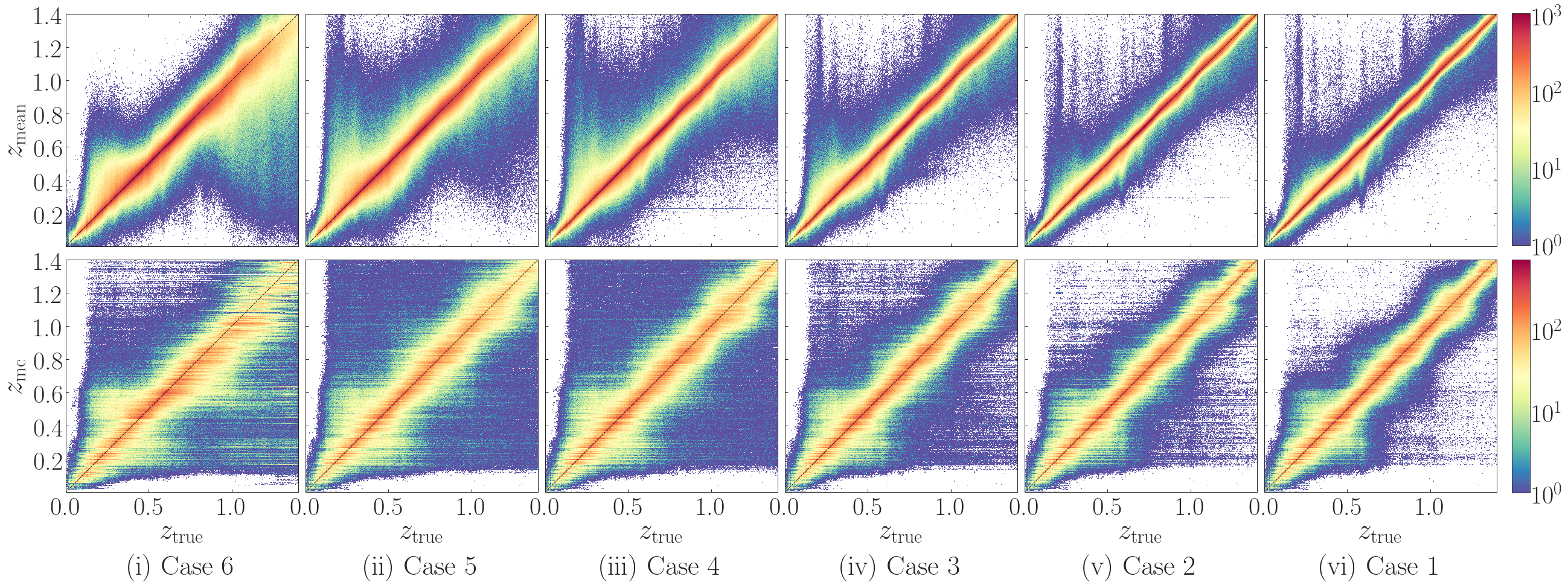}
    \caption[]{Scatter plot of both photometric redshifts given by DNF, $z_{\rm mean}$ (top row) and $z_{\rm mc}$ (bottom row), as a function of true redshift for all the samples described in \Cref{sec:Samples} up to \VIS $< 24.5$. The $\sigma$ of photo-$z$ for these sample at \VIS $< 24.5$ is from left to right: 0.063, 0.049, 0.046, 0.036, 0.032, 0.029.
     }
    \label{fig:zmean_vs_true_redshift_vis_245}
\end{figure*}

\begin{table*}
	\centering
	\caption{Photo-$z$ metrics of each photometric sample and cut in \VIS (as explained in \Cref{sec:Dep_photoz_sample_size}). 
	}
	\label{tab:Photoz_stats_table}
    \begin{tabular}{lcccccc}
    \noalign{\smallskip}
    \multicolumn{7}{c}{\textbf{Normalised Median Absolute Deviation}}  \\ 
	\noalign{\smallskip}
	\hline
	\hline
	\noalign{\smallskip}
    \VIS &  Shallow depth inho. &  Shallow depth &  Mid depth \Euclid &  Mid depth &  Fiducial &  Optimistic \\
   	\noalign{\smallskip}
	\hline
	\noalign{\smallskip}
    25   &                0.090 &          0.066 &             0.061 &      0.046 &     0.040 &       0.036 \\
    24.5 &                0.063 &          0.049 &             0.046 &      0.036 &     0.032 &       0.029 \\
    24   &                0.049 &          0.039 &             0.038 &      0.031 &     0.028 &       0.026 \\
    23.5 &                0.041 &          0.033 &             0.034 &      0.027 &     0.025 &       0.024 \\
    23   &                0.036 &          0.029 &             0.030 &      0.024 &     0.023 &       0.022 \\
	   	\noalign{\smallskip}
		\hline
	    \noalign{\smallskip}
    \multicolumn{7}{c}{\textbf{Fraction of outliers (\%)}}  \\ 
    	\noalign{\smallskip}
		\hline
	    \noalign{\smallskip}
	\noalign{\smallskip}
    25   &                 25.8 &           16.1 &              14.4 &        9.0 &       6.9 &         5.1 \\
    24.5 &                 12.9 &            7.5 &               6.3 &        3.3 &       2.2 &         1.5 \\
    24   &                  5.5 &            3.6 &               3.0 &        1.6 &       1.0 &         0.8 \\
    23.5 &                  2.8 &            1.9 &               1.7 &        0.8 &       0.6 &         0.5 \\
    23   &                  1.6 &            1.0 &               0.9 &        0.4 &       0.3 &         0.3 \\
		\noalign{\smallskip}
		\hline
		\hline
	    \noalign{\smallskip}
	\end{tabular}
\end{table*}

The cosmological tomographic analysis of a photometric survey divides the whole sample into redshift bins selected with a photo-$z$ technique. In our study, we want to follow as close as possible the methodological steps that one would carry out in real surveys. For that purpose, we compute the photo-$z$s of all our study cases described in \Cref{tab:cases}. We use the Directional Neighborhood Fitting (DNF; \citealt{Vicente_2016}) training-based algorithm to estimate realistic photo-$z$ estimates of our simulated galaxies. The exact choice of the machine learning training set method is not important for our analysis as most methods of this type perform similarly to the precision levels we are interested in \citep[see e.g.][]{Euclid_Collaboration_2020, Sanchez_2014}.

DNF estimates the photo-$z$ of a galaxy based on its closeness in observable space 
to a set of training galaxies whose redshifts are known. The main feature of DNF is that the metric that defines the distance or closeness between objects is given by a directional neighbourhood metric, which is the product of a Euclidean and an angular neighbourhood metrics. This metric ensures that neighbouring objects are close in colour and magnitude space. The algorithm fits a linear adjustment, a hyperplane, to the directional neighbourhood of a galaxy to get an estimation of the photo-$z$. This photo-$z$ estimate is called $z_{\rm mean}$, which is the average of the redshifts from the neighbourhood. The residual of the fit is considered as the estimation of the photo-$z$ error. In addition, 
DNF also produces another photometric redshift estimate, $z_{\rm mc}$ that is a Monte Carlo draw from the nearest neighbour in the DNF metric for each object. 
Therefore, it can be considered as a one-point sampling of the photo-$z$ probability density distribution. As such, it is not a good individual photo-$z$ estimate of the object, but when all the estimates in a galaxy sample are stacked it can recover the overall probability density distribution of the sample~\citep{Rau_2016}. When working with tomographic bins, we will classify the galaxies into different bins using their $z_{\rm mean}$ and we will obtain the photometric distribution, $n(z)$, within each bin by stacking their $z_{\rm mc}$.
This is an approach used by DES in analysing their First Year Data results (e.g., Hoyle et al. 2018, Crocce et al. 2019, Camacho et al. 2019) providing redshift distributions that were validated with other independent assessment methods. Therefore, we define the $n(z)$ by stacking the $z_{\rm mc}$ estimator instead of the true redshift of the simulation to make the photo-$z$ distribution close to what would be obtained in a real data analysis with the assurance that the method has been validated. 

We select a patch of sky of $3.35$ deg$^{2}$ to create the samples to train DNF. These training samples have the magnitudes and errors computed with the same magnitude limits as the sample whose photo-$z$ we want to compute (see \Cref{tab:Coadded_depths_10sigma}). We generate three types of spectroscopic training samples. For all of them we limit the spectroscopic training sample to galaxies brighter than \VIS $=24.5$ as there are few objects whose redshift has been reliably determined beyond that magnitude limit. 
The spectroscopic training samples are described in \cref{sec:Samples}.

The true redshift distributions of the spectroscopic training set used to train DNF for each of the sample cases considered here are shown in \Cref{fig:samples_training_redshift_distributions}.
In blue, we present the redshift distribution of case 1 with the first spectroscopic training sample that it is fully complete as a function of magnitude. We show in black the resulting $N(z)$ of case 2. Cases 3--5 (olive, red and orange colours in \Cref{fig:samples_training_redshift_distributions,fig:samples_redshift_distributions}) have the same training sample completeness as a function of magnitude. The drop in completeness at faint magnitudes translates into a decrease of objects at high redshift. Last, we present the resulting redshift distribution with the third spectroscopic training set in orange. Gathering multiple selection criteria from different spectroscopic surveys leads to an inhomogeneous redshift distribution for the spectroscopic training sample. In \Cref{fig:samples_redshift_distributions}, we show the overall photo-$z$ distributions of $z_{\rm mean}$ (top panel) and $z_{\rm mc}$ (bottom panel) values obtained for the full sample for each of the six cases. We see how an inhomogeneous $N(z)$ in the training sample leads to an inhomogeneous distribution of the photo-$z$.

\Cref{fig:zmean_vs_true_redshift_vis_245} shows the photo-$z$ obtained with DNF as a function of true redshift for the six samples up to \VIS $< 24.5$. This figure gives us an indication of how the photo-$z$ scatter decreases with deeper photometry. Photometric samples go up to \VIS $=25$. However, we cut the spectroscopic training sample at \VIS $=24.5$ to be more realistic. The lack of objects between $24.5$ and $25.0$ in the training sample forces the algorithm to extrapolate beyond that magnitude, and thus noisier photometric redshifts are obtained. In \Cref{fig:zmean_vs_true_redshift_vis_245}, we show galaxies only down to \VIS $< 24.5$ to reduce the noise and make the figure clearer.

To quantify the photo-$z$ precision for the different samples we use the following typical metrics:
\begin{itemize}
    \item The normalised median absolute deviation:
    \begin{equation}
    \sigma=1.4826\cdot\textrm{median}(|\Delta z-\textrm{median}(\Delta z)|)\,,
    \end{equation}
    where 
    \begin{equation}
    \Delta z= \frac{z_{\textrm{spec}}-z_{\textrm{phot}}}{1+z_{\textrm{spec}}}\,.
    \end{equation}
    \item We consider outliers those objects with $|\Delta z|>0.15$.
\end{itemize}
In \Cref{tab:Photoz_stats_table} we show the values obtained for these metrics for each photometric sample.



\section{Building forecasts for \Euclid}\label{sec:building-forecasts}

So far, we have seen how the photometric depth and the spectroscopic training sample determine the overall redshift distributions of the resulting samples. We have selected six cases to cover a range of possible scenarios that we may encounter in the analysis of \Euclid data complemented with ground-based surveys. Once the galaxy distributions for the photometric cases under study have been obtained, we want to propagate the photo-$z$ accuracy in determining tomographic subsamples to the final constraints on the cosmological parameters in order to understand how to optimize the photometric sample for galaxy clustering analyses.

We follow the forecasting prescription presented in \citet[][hereafter EC20]{IST:paper1}. We consider the same Fisher matrix formalism and make use of the \texttt{CosmoSIS}\footnote{\url{https://bitbucket.org/joezuntz/cosmosis/wiki/Home}} code validated for \Euclid specifications therein. Our observable is the tomographically binned projected angular power spectrum, $C_{ij}(\ell)$, where $\ell$ denotes the angular multipole, and $i,j$ stand for pairs of tomographic redshift bins. This formalism is the same for WL, galaxy clustering (with the photometric sample), and GGL, with the only difference being the kernels used in the projection from the power spectrum of matter perturbations to the spherical harmonic-space observable. 
We focus on the \GCph cosmological probe, as well as its combination with GGL. The projection to $C_{ij}(\ell)$ is performed under the Limber, flat-sky and spatially flat approximations \citep{2017MNRAS.469.2737K,2017MNRAS.472.2126K, 2018PhRvD..98b3522T}. We also ignore redshift-space distortions, magnification, and other relativistic effects \citep{2019arXiv191207326D}. To minimise the impact of neglecting relativistic effects, more relevant at large scales, in our analysis we consider multipole scales from $\ell\geq10$ to $\ell\leq750$, which corresponds to the more conservative scenario in \citetalias{IST:paper1}.

\begin{figure*}
    \includegraphics[width=\textwidth]{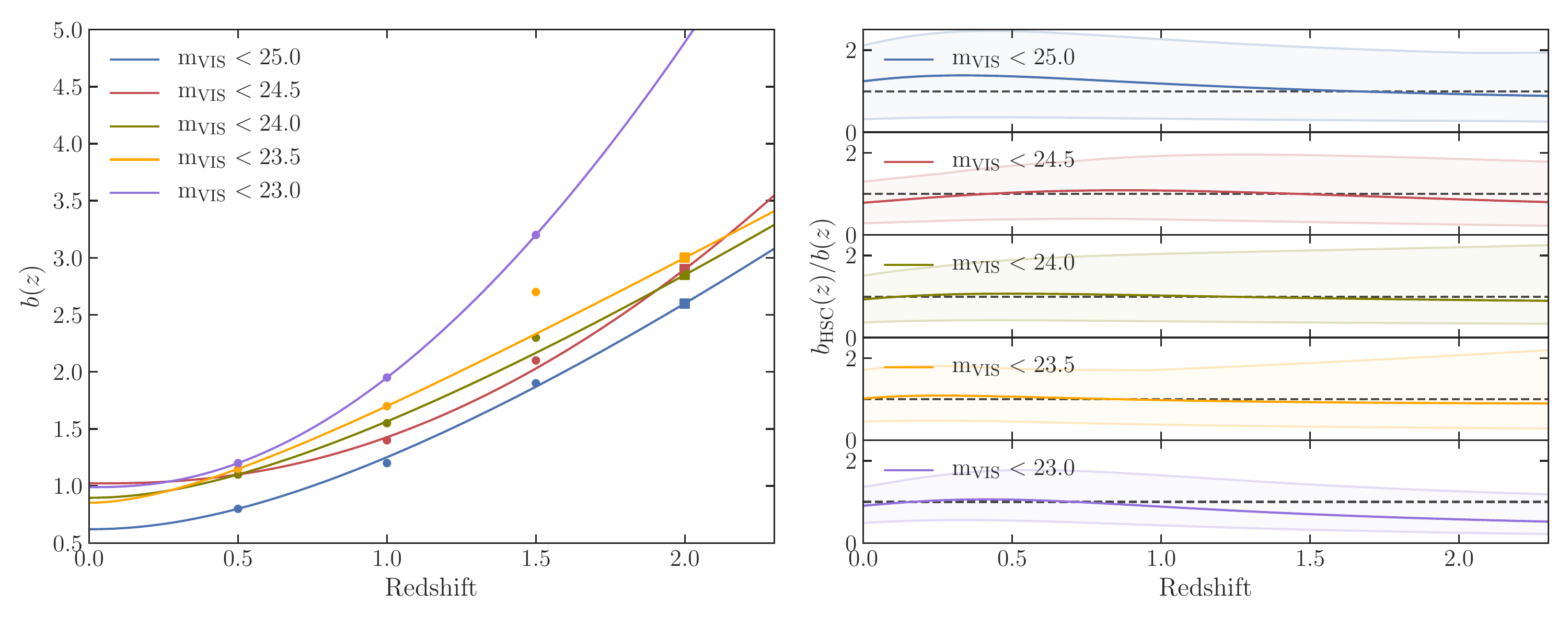}
    \caption{\textit{Left panel:} Galaxy bias as a function of redshift. Dots correspond to the measured values in the Flagship simulation for different magnitude cuts and the solid lines are a fit following \Cref{eq:Flagship_bias}. We plot with squares the bias values obtained for $z = 2$ to indicate that at that redshift there are few objects and thus the values are slightly less reliable. At \VIS $<23$ there were not enough objects at $z = 2$ to compute the bias in Flagship. \textit{Right panel:} Ratio between the HSC bias, $b_{\rm HSC}$, from N20 and the Flagship bias for each magnitude-limited sample.
    To assess the $1\sigma$ uncertainty of $b_{\rm HSC}$ along the redshift range, we generate a set of Gaussian random numbers for the free parameter $\alpha$, $b_{1}$, and $b_{0}$ of $b_{\rm HSC}$ with their values as mean and their errors as standard deviation. Then we evaluate $b_{\rm HSC}$ in the redshift range for all the set of free parameters previously generated. We pick the maximum and minimum $b_{\rm HSC}$ at each redshift. This corresponds to the shaded regions.}
    \label{fig:bias}
\end{figure*}

When considering GGL, its power spectrum contains contributions from galaxy clustering and cosmic shear, but also from intrinsic galaxy alignments (IA). We assume the latter is caused by a change in galaxy ellipticity that is linear in the density field. Note that such modelling is appropriate for large scales \citep{DESY1-WL}, like the ones considered in this analysis, but more complex models should be used for the very small scales \citep[see e.g.][]{Blazek2019,MAFortuna}. Under this linear assumption we can define the density-intrinsic and intrinsic-intrinsic three-dimensional power spectra, $P_{\delta \rm I}$ and $P_{\rm II}$, respectively. They can be related to the density power spectrum $P_{\delta\delta}$ with $P_{\delta \rm I}=-A(z)P_{\delta\delta}$ and $P_{\rm II}=A(z)^2P_{\delta\delta}$. We follow \citetalias{IST:paper1} in parameterising $A$ as
\begin{equation}
    A(z)=\frac{\mathcal{A}_{\rm IA}\mathcal{C}_{\rm IA}\Omega_{\rm m}\mathcal{F}_{\rm IA}(z)}{D(z)}\,,
\end{equation}
where $\mathcal{C}_{\rm IA}$ is a normalisation parameter that we set as $\mathcal{C}_{\rm IA}=0.0134$, $D(z)$ is the growth factor, and $\mathcal{A}_{\rm IA}$ is a nuisance parameter fixing the amplitude of the IA contribution.

We model the redshift dependence of the IA contribution as
\begin{equation}
    \mathcal{F}_{\rm IA}=(1+z)^{\eta_{\rm IA}}\left[\frac{\braket{L}(z)}{L_*(z)}\right]^{\beta_{\rm IA}}\,,
\end{equation}
with $\braket{L}(z)/L_*(z)$ being the redshift-dependent ratio between the average source luminosity and the characteristic scale of the luminosity function \citep{Hirata_2007, Bridle_2007}. For a detailed explanation on IA modeling see \citet{Samuroff_2019}.
We use the same ratio of luminosities for every galaxy sample. However, this ratio should in principle depend on the specific galaxy population. Since we select galaxies according to a \VIS cut and not according to a particular galaxy type, we expect that the luminosity ratio does not change significantly between galaxy samples, and therefore use the same ratio for simplicity.
We set the fiducial values for the intrinsic alignments nuisance parameters to
\begin{equation}
    \{\mathcal{A}_{\rm IA},\eta_{\rm IA},\beta_{\rm IA}\}=\{1.72,-0.41,2.17\}\,,
\end{equation}
in agreement with the recent fit to the IA contribution in the Horizon-AGN simulation \citep{2015MNRAS.454.2736C}, although the amplitude $\mathcal{A}_{\rm IA}$ might be smaller in practice \citep{MAFortuna}.

When considering \GCph and GGL one of the primary sources of uncertainty is the relation between the galaxy distribution and the underlying total matter distribution, i.e. the galaxy bias \citep{1987MNRAS.227....1K}. 
We consider a linear galaxy bias relating the galaxy density fluctuation to the matter density fluctuation with a simple linear relation
\begin{equation}
    \delta_{\rm g}(\vec{x}, z) = b(z) \delta_{\rm m}(\vec{x}, z)\,, 
\end{equation}
where we neglect any possible scale dependence. Note that a linear bias approximation is sufficiently accurate for large scales \citep{DESY1-GCWL}. However, when adding very small scales into the analysis, a more detailed modeling of the galaxy bias is required \citep[see e.g.,][]{2017MNRAS.464.1640S}. One of the approaches to this modeling is through perturbation theory, which introduces a nonlinear and nonlocal galaxy bias \citep{2018PhR...733....1D}. 

We consider a constant galaxy bias in each tomographic bin. We get their fiducial values by fitting the directly measured bias in Flagship to the function
\begin{equation}\label{eq:Flagship_bias}
    b(z) = \frac{Az^{B}}{1+z}+C\,, 
\end{equation}
where $A$, $B$ and $C$ are nuisance parameters. 
We select five subsamples with \VIS limiting magnitudes: $25$, $24.5$, $24$, $23.5$, and $23$ from the Flagship galaxy sample. We compute the bias values as a function of redshift for each of these magnitude-limited subsamples using directly the true redshift of Flagship at redshifts 0.5, 1, 1.5 and 2. 
As an approximation, we use the same galaxy bias for each of the six photometric samples and change the fiducial according to the magnitude limit cut.
The obtained bias and fitted functions are shown in the left panel of \Cref{fig:bias}. To fit the bias-redshift relation we choose to use all galaxy bias values computed with the Flagship simulation, although values at $z = 2$ were less reliable. 
The value of the bias at $z = 1.5$ falls outside the bias-redshift fit
for the \VIS $<23$ sample. However, we recomputed the bias fit neglecting the value at $z=2$ and including the value at $z=1.5$, but no significant changes were appreciated, therefore we keep the bias computed using the fits shown in \Cref{fig:bias}.

To validate the bias obtained with Flagship, we compare our bias values to the ones obtained from the Hyper Suprime-Cam Subaru Strategic Program (HSC-SSP) data release 1 (DR1) by \citet[][N20 hereafter]{Nicola_2020}. The HSC survey has comparable survey depth and uses similar ground-based bands to the ones considered in this work. N20 fit galaxy bias on magnitude-limited galaxy samples down to $i < 24.5$. We compare their values to ours in the right panel of~\Cref{fig:bias}. We extrapolate their bias down to $i < 25$ for 
our faintest magnitude bin. Strictly speaking, we are comparing $i$-band magnitude-selected samples from N20 to our \VIS-band magnitude-selected samples. We have checked in Flagship that the bias values for both $i$-band and \VIS-band selected samples cut at the same magnitude limit do not change by more than 10\% and therefore our comparison is meaningful. N20 assume that bias can be split into two separated terms of redshift and limiting magnitude, and define it as
\begin{equation}
    b_{\rm HSC}(z, m_{\rm lim}) = \bar{b}(m_{\rm lim})D^{\alpha}(z)\,,
\end{equation}
where $\alpha$ is a variable that takes into account the inverse relation between the growth factor and galaxy bias. By fitting $\alpha$ and $\bar{b}(m_{\rm lim})$ in a multi-step weighted process they find
\begin{equation}
    \begin{split}
    \alpha & = -1.30\pm 0.19\,,\\
    \bar{b}(m_{\rm lim}) & = b_{1}(m_{\rm lim}-24)+b_{0}\,,
    \end{split}
\end{equation}
where $b_{1} = -0.0624 \pm 0.0070$ and $b_{0} = 0.8346 \pm 0.161$. For a detailed explanation see Sect. 4.6 in N20. We compute $D(z)$ for our sample and use our \VIS magnitude cuts as $m_{\rm lim}$ along with their fitted parameters to get a bias to compare. The ratio between the HSC bias, $b_{\rm HSC}$, and ours, $b(z)$, is shown in the right panel of \Cref{fig:bias}. Note that N20 compute their bias up to redshift 1.25 and that we have extrapolated their behaviour to higher redshifts for the comparison at $z>1.25$. The values of the bias in Flagship stay within 1$\sigma$ of the HSC values, $b_{\rm HSC}$ (shaded area in the right panel of \Cref{fig:bias}), confirming that the bias values we use are consistent with the HSC observations.

We consider the same redshift distributions for both \GCph and GGL. In practice, this is an over-simplification, since these two probes will probably apply different selection criteria when determining their samples. GGL for instance will give some importance to the shape measurements of the galaxies. But for the present Fisher matrix analysis we limit ourselves to use the same sample for both probes.



\section{Cosmological model}\label{sec:cosmo}

We optimize the photometric sample of \Euclid considering the baseline cosmological model presented in \citetalias{IST:paper1}: a spatially flat Universe filled with cold dark matter and dark energy. We approximate the dark energy equation of state parameter with the CPL \citep{Chevallier_Polarski_2001,2005PhRvD..72d3529L} parameterisation
\begin{equation}
    w(z)=w_0+w_a\frac{z}{1+z}\,.
\end{equation}

Therefore, the cosmological model is fully specified by the dark energy parameters, $w_0$ and $w_a$, the total matter and baryon density today, $\Omega_{\rm m}$ and $\Omega_{\rm b}$, the dimensionless Hubble constant, $h$, the spectral index, $n_{\rm s}$, and the RMS of matter fluctuations on spheres of 8 $h^{-1}$Mpc radius, $\sigma_8$. We assume a dynamically evolving, minimally-coupled scalar field, with sound speed equal to the speed of light and vanishing anisotropic stress as dark energy. Therefore, we neglect any dark energy perturbations in our analysis. We also allow the equation of state of dark energy to cross $w(z)=-1$ using the \citet{Hu:2007pj} prescription.

The fiducial values of the cosmological parameters are given by
\begin{align}
    \{\Omega_{\rm m},\Omega_{\rm b}, w_0, w_a, \mathrm{h}&, n_{\rm s}, \sigma_8\}=\nonumber\\
    &=\{0.32,0.05,-1,0,0.67,0.96,0.816\}\,.
\end{align}

Moreover, we fix the sum of neutrino masses to $\sum m_{\nu}=0.06\,$eV. Note that the linear growth factor depends on both redshift and scale when neutrinos are massive, but we follow \citetalias{IST:paper1} in neglecting this effect, given the small fiducial value considered. Therefore, we compute the growth factor accounting for massive neutrinos, but neglect any scale dependence. Note that the fiducial values used in this analysis are compatible with the fiducial cosmology of the Flagship simulation presented in \Cref{sec:Flagship} except for $\sigma_8$. This can be explained by the fact that the Flagship simulation does not account for massive neutrinos and therefore considers a slightly larger value for $\sigma_8$. However, since we are only extracting the galaxy bias and the galaxy distributions from Flagship, and we are computing Fisher forecasts, this difference in the fiducial $\sigma_8$ value does not have any impact on our results.

We quantify the performance of photometric galaxy samples in constraining cosmological parameters through the metric figure of merit (FoM), as defined in \citet{Albrecht_2006} but with the parameterisation defined in \citetalias{IST:paper1}. 
Our FoM is proportional to the inverse of the area 
of the error ellipse in the parameter plane of $w_0$ and $w_a$ defined by the marginalised Fisher submatrix, $\mathbf{\Tilde{F}}_{w_{0}w_{a}}$,
\begin{equation}\label{eq:FoM}
    \textrm{FoM}_{w_{0}w_{a}} = \sqrt{\textrm{det}\left(\mathbf{\Tilde{F}}_{w_{0}w_{a}}\right)}\,.
\end{equation}
We will use the FoM defined above throughout this article. The higher the FoM value, the higher the cosmological constraining power.


\section{Results}\label{sec:results}

In this section we carry out a series of tests to optimize the sample selection for \GCph analyses. We want to determine the best number and type of tomographic bins to constrain cosmological parameters. We explore the influence of the accuracy in the photo-$z$ estimation and sample size in providing cosmological constraints. 
We split the data in tomographic redshift bins in order to have more control in the variations of sample size and photo-$z$ accuracy to better understand their impact in constraining cosmological parameters.
We use the FoM defined in \Cref{eq:FoM} to quantify the constraining power on the cosmological parameters. In addition, we also compute the FoM when combining \GCph  with GGL, assuming the same photo-$z$ sample, which implies the same photo-$z$ binning and number density. When computing the cosmological constraining power for \GCph + GGL, we marginalize over the galaxy bias of each tomographic bin and intrinsic alignment parameters, whereas for \GCph alone the galaxy bias parameters are fixed to their fiducial values. The main reason for this choice is that, under the linear galaxy bias approximation, there is a large degeneracy between the galaxy bias and $\sigma_8$. In this case, the Gaussianity assumption of the Fisher matrix approach breaks down and its constraints on the cosmological parameters are not reliable. Therefore, we fix the galaxy bias to break this degeneracy when considering \GCph alone. Note that when we combine \GCph with GGL, the additional information brought by the latter is enough to break such degeneracy and constrain $\sigma_8$ and the galaxy bias at the same time.

\subsection{Optimizing the type and number of tomographic bins}

\begin{figure}
    \includegraphics[width=\columnwidth]{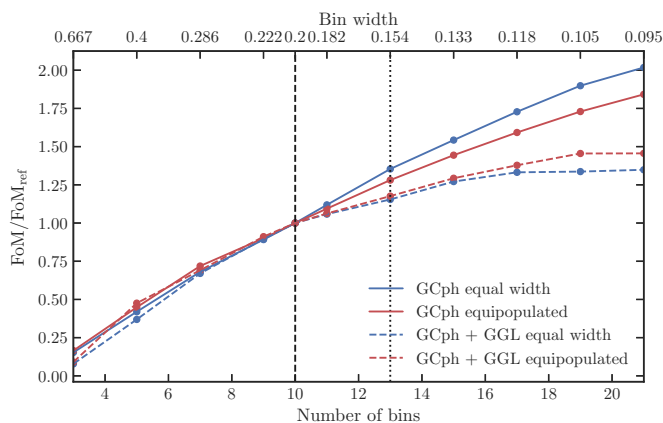}
    \caption{FoM as a function of the number of bins for \GCph (solid) and \GCph+\ GGL (dashed) and for bins with the same redshift width (blue) and with the same number of objects (red). The redshift width of the bins when they have the same width is shown in the top x axis. The FoMs are normalised to the FoM at 10 bins, FoM$_{\text{ref}}$, which corresponds to the specifications for the number of bins used to compute the forecasts in \citetalias{IST:paper1} and denoted by a vertical-dashed line. A vertical-dotted line shows the 13 bins used as our fiducial choice. 
    }
    \label{fig:FoM_vs_nbins}
\end{figure}

We bin galaxies into different numbers of redshift bins to study the impact of the number of redshift bins on the cosmological parameter inference. When we define redshifts bins, we choose galaxies within the redshift range $[0, 2]$ since the maximum lightcone outputs generated in 
Flagship are at $z = 2.3$ and we prefer to avoid working at the limit of the simulation. We check the effect of using bins with the same redshift width and bins with the same number of objects (equipopulated). We also see the difference when using only \GCph or both \GCph and GGL probes. This analysis is performed using our fiducial sample (case 2) up to \VIS $<24.5$. We compute the FoM for all the cases mentioned and show the results in \Cref{fig:FoM_vs_nbins}. The FoM are normalised to ten bins since this is the default number used to compute the forecasts in \citetalias{IST:paper1}.

As seen in \Cref{fig:FoM_vs_nbins}, the general tendency of the FoM is to increase with the number of bins. 
\citetalias{IST:paper1} used ten tomographic bins as their fiducial value.
For bins with equal width in redshift, the FoM increase when moving from ten to thirteen bins is $35.4\%$ and $15.4\%$, for \GCph only and for \GCph + GGL, respectively. The FoM improvement we get from going to even more bins does not compensate the increase in computational time needed for the analysis. This is especially true when using both probes, where we notice that the curve flattens while in \GCph the FoM continues to increase since the bias is fixed and thus the amount of information that can be extracted is larger than expected in practise. Moreover, our photo-z treatment may start to be too simplistic to realistically deal with too many photometric redshift bins.  
The FoM saturates with the increasing number of bins because it is not possible to extract more information on radial clustering when the width of the bins is smaller than the photo-$z$ precision. At this limit, the uncertainty at which bin a particular galaxy belongs is greatly increased. For \GCph + GGL the curves flatten at lower number of bins since systematic effects in the marginalisation of galaxy bias and intrinsic alignment free parameters also affect the cosmological information that can be extracted. Therefore, we choose thirteen to be our fiducial number of bins as a conservative choice. 

In addition, we choose bins with equal width in redshift as the optimal way of partitioning the sample since we observe that, overall, for \GCph the FoM is larger in this case than in the equipopulated one.
For thirteen bins with equal width the FoM is $713$ while it is $547$ for equipopulated bins\,\footnote{Recall that galaxy bias is fixed when considering \GCph alone, which provides these large absolute values for the FoMs.}, which is an increase of $30\%$. For the \GCph + GGL combined analysis, the FoM does not appreciably change between the use of bins with equal width and equipopulated ones. At thirteen bins, which is the fiducial choice, the FoM difference of using bins with equal width or equipopulated ones is negligible.

\begin{figure}
	\includegraphics[width=\columnwidth]{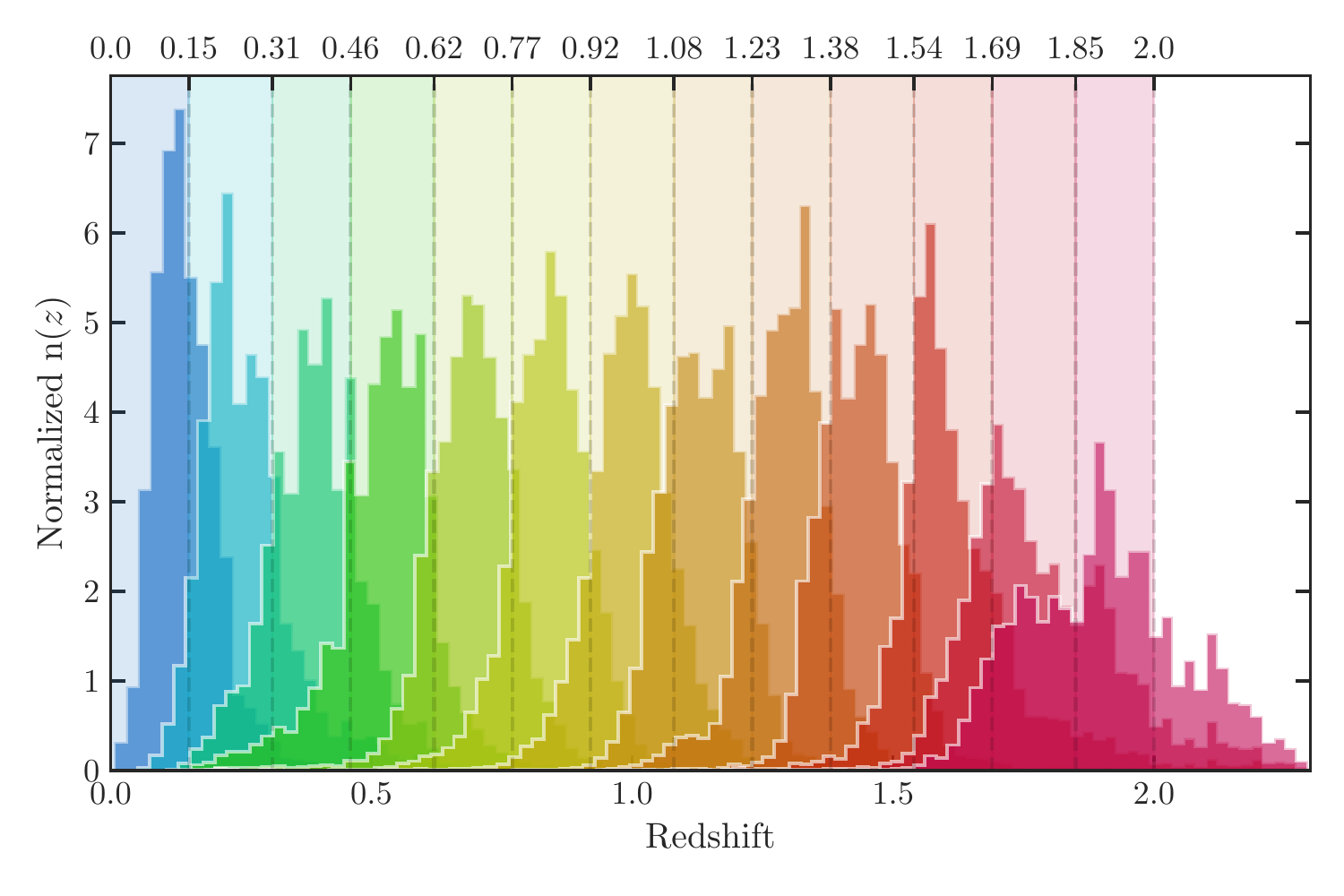}
    \caption{Redshift distributions ($z_{\rm mc}$) of the thirteen bins with equal width for the fiducial sample (case 2). 
    The top $x$-axis correspond to the values of the redshift limits of the thirteen bins with equal width in redshift.
    The shaded regions indicate these limits.}
    \label{fig:Nofz_13_bins_fiducial_sample}
\end{figure}
 
We will use these bin choices to analyze the dependency of cosmological constraints on the photo-$z$ quality and size of the sample. In \Cref{fig:Nofz_13_bins_fiducial_sample} we show the redshift distributions of the thirteen bins with equal width for our fiducial case 2 sample.

\subsection{FoM dependency on photometric redshift quality and number density}\label{sec:Dep_photoz_sample_size}

\begin{figure*}[ht]
	\includegraphics[width=\textwidth]{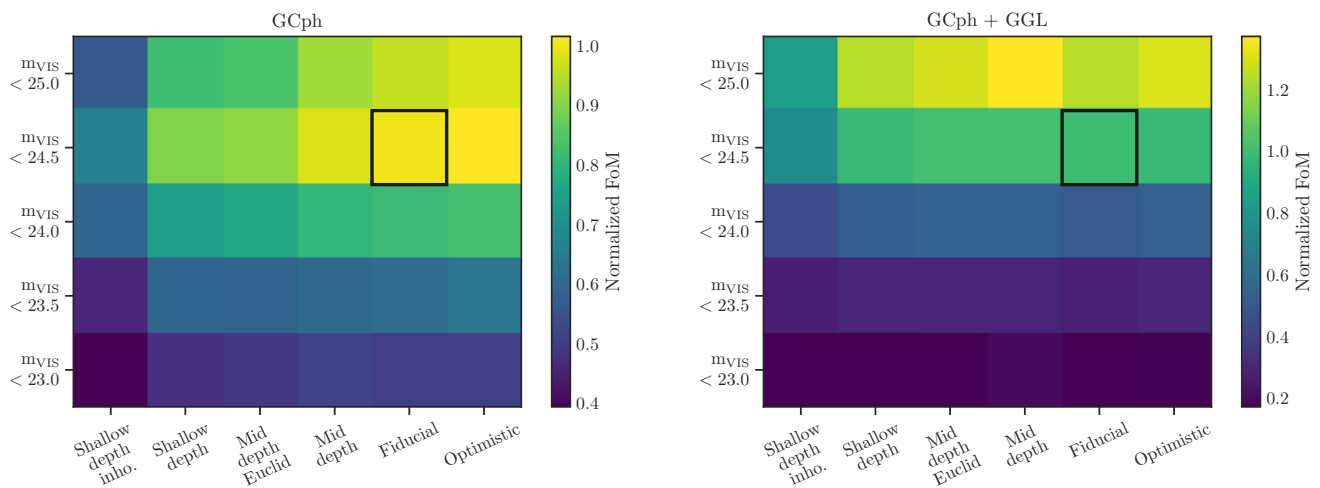}
    \caption{FoM for the samples defined in \Cref{sec:Samples} with different photo-$z$ accuracy and sample size. The size has been reduced by performing a series of cuts in \VIS. The results are normalised to the FoM of the fiducial sample with \VIS $<24.5$ (highlighted cell). The figures correspond to the results for using only \GCph (left) and for combining it with GGL (right).}
    \label{fig:FoM_vs_mag_pz}
\end{figure*}

\begin{table*}[ht]
	\centering
	\caption{Values of the FoM for samples defined in \Cref{sec:Samples} with different photo-$z$ accuracy and sample size (same cases as in \Cref{fig:FoM_vs_mag_pz}). 
	The results are normalised to the FoM of the fiducial sample with \VIS $<24.5$. For reference, the unnormalised value of our fiducial sample is $713$ for \GCph and $411$ for \GCph + GGL. Note that galaxy bias and intrinsic alignments nuisance parameters are free in the latter, which provides a lower FoM than in \GCph alone.}
	\label{tab:FoM_vs_mag_pz_table}
    \begin{tabular}{lcccccc}
    \noalign{\smallskip}
    \multicolumn{7}{c}{\textbf{\GCph}}  \\ 
	\noalign{\smallskip}
	\hline
	\hline
	\noalign{\smallskip}
    \VIS &  Shallow depth inho. &  Shallow depth &  Mid depth \Euclid &  Mid depth &  Fiducial &  Optimistic \\
   	\noalign{\smallskip}
	\hline
	\noalign{\smallskip}
    25   &                0.57 &          0.82 &              0.84 &       0.93 &      0.96 &        0.98 \\
    24.5 &                0.67 &          0.90 &              0.91 &       0.98 &      1.00 &        1.02 \\
    24   &                0.59 &          0.74 &              0.77 &       0.81 &      0.82 &        0.83 \\
    23.5 &                0.46 &          0.59 &              0.59 &       0.61 &      0.62 &        0.64 \\
    23   &                0.39 &          0.48 &              0.50 &       0.52 &      0.51 &        0.51 \\
	   	\noalign{\smallskip}
		\hline
	    \noalign{\smallskip}
    \multicolumn{7}{c}{\textbf{\GCph and GGL}}  \\ 
    	\noalign{\smallskip}
		\hline
	    \noalign{\smallskip}
	\noalign{\smallskip}
    25   &                0.85 &          1.24 &              1.29 &       1.37 &      1.24 &        1.30 \\
    24.5 &                0.75 &          0.98 &              1.01 &       1.01 &      1.00 &        0.98 \\
    24   &                0.46 &          0.53 &              0.55 &       0.54 &      0.52 &        0.54 \\
    23.5 &                0.27 &          0.30 &              0.30 &       0.29 &      0.28 &        0.30 \\
    23   &                0.17 &          0.17 &              0.18 &       0.20 &      0.17 &        0.18 \\
		\noalign{\smallskip}
		\hline
		\hline
	    \noalign{\smallskip}
	\end{tabular}
\end{table*}

Another aspect we want to study is the effect of the trade-off between photo-$z$ accuracy and number density on the constraining power of cosmological parameters. For that purpose, we take the six photometric samples defined in \Cref{sec:Samples} and apply five cuts ($25$, $24.5$, $24$, $23.5$, $23$) in \VIS to modify the sample size (leading to a number density of about 41, 29, 18, 12 and 9 galaxies per arcmin$^{2}$ respectively). Besides reducing the number density of the photometric samples, the cut in \VIS also affects the photo-$z$ distribution and accuracy of the overall sample. A bright magnitude cut, that eliminates the fainter galaxies, mostly removes galaxies with higher and thus less reliable redshifts.
We compute the FoM for all the cases mentioned before and normalise them to the FoM of our fiducial (case 2) sample at \VIS $< 24.5$, for both \GCph only and \GCph + GGL. To help visualise the results, we present the resulting FoM in a grid format in \Cref{fig:FoM_vs_mag_pz} and the values themselves in \Cref{tab:FoM_vs_mag_pz_table}.
The configuration of tomographic bins used to perform the analysis is the optimum one found in the previous section, which is thirteen bins with equal width.

Let us first discuss the case of \GCph alone. As seen in \Cref{fig:FoM_vs_mag_pz}, in general, the FoM for \GCph increases with deeper photometric data, which improves the photo-$z$ performance (increasing along the $x$-axis in the figure). The FoM also increases with number density, determined by the magnitude limit imposed (increasing along the $y$-axis). 
We notice a larger increase in the FoM with sample size in those samples where the photo-$z$ quality is better (e.g., the optimistic, fiducial and mid depth ground-based photometry cases). In these cases, increasing the sample size from a \VIS cut from 23.5 to 24 and from 24 to 24.5 leads to an increase of the FoM of about $20\%$. Clearly, having a fainter magnitude cut results in larger samples that yield higher FoM values. This trend is in agreement with the results presented in \citet{Tanoglidis_2019}.


\begin{figure}
	\includegraphics[width=\columnwidth]{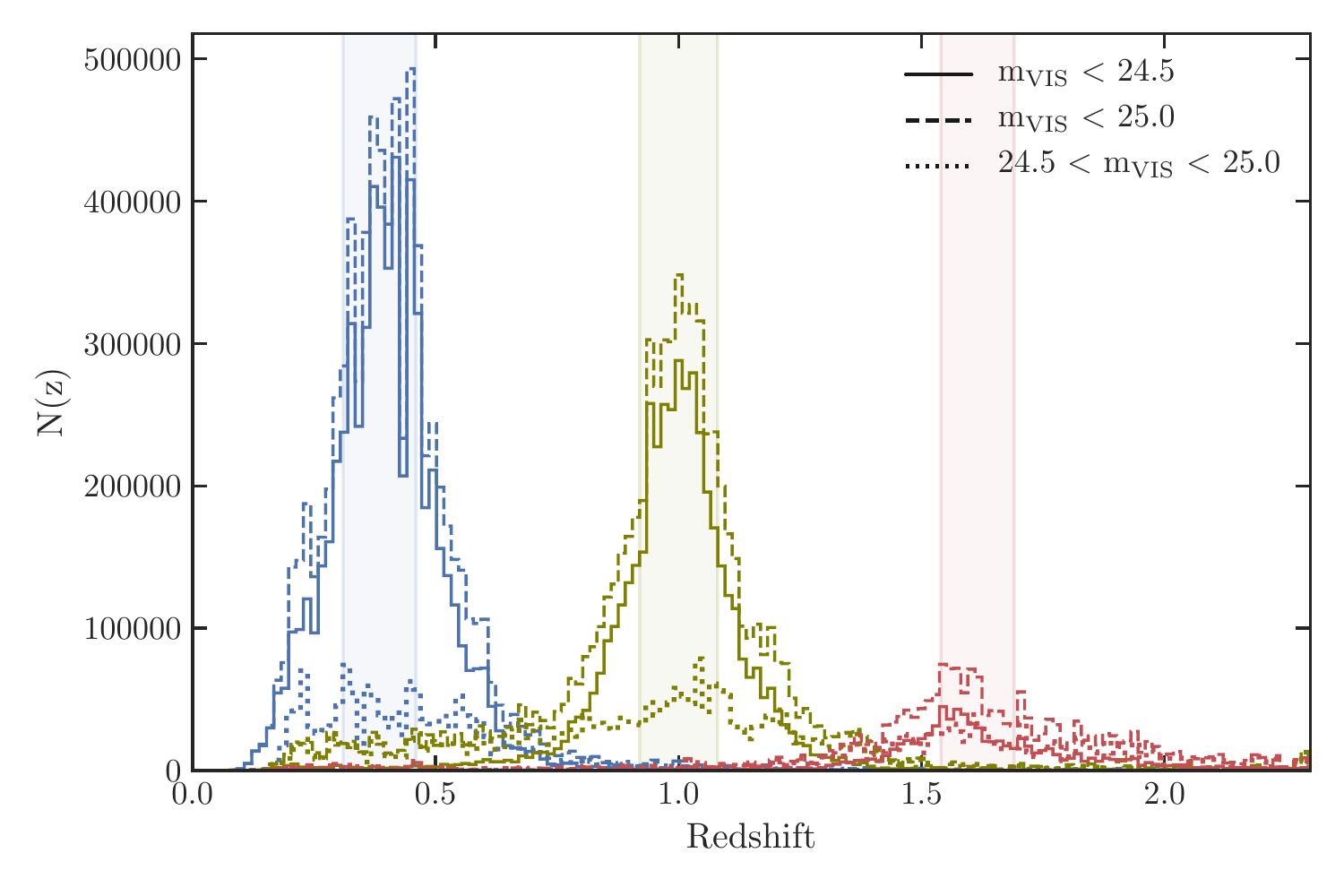}
    \caption{Redshift distributions ($z_{\rm mc}$) of three of the thirteen tomographic bins selected with $z_{\rm mean}$ corresponding to the shaded regions $[0.31, 0.46)$ (blue), $[0.92, 1.08)$ (green) and $[1.54, 1.69)$ (red). For each bin we plot objects with \VIS $<24.5$ (solid), \VIS $<25.0$ (dashed) and $24.5<$ \VIS $<25.0$ (dotted). The photometric sample used is the mid depth \Euclid (case 4).}
    \label{fig:Nofz_3_bins_VIS_cuts}
\end{figure}

The trend of increasing FoM as we take fainter magnitude limit cuts and increase the number density continues as long as the photo-$z$ performance is not degraded. Once we push to faint magnitudes where there are no objects to train the photometric redshift algorithms, their performance degrades and the photometric redshift bins start to be wider. 
There are many object that do not belong to the bins and spurious cross-correlations between different bins appear. As a result, the strength of the cosmological signal is diminished and the FoM decreases. This effect can be seen in \Cref{fig:FoM_vs_mag_pz} for the \GCph case (left panel), where we can appreciate a reduction in the FoM when we move from a magnitude-limited sample cut at \VIS $<24.5$ (second row from the top) to a magnitude-limited sample cut at \VIS $<25.0$ (top row). With this change, we are increasing the sample, but with galaxies that cannot be located in redshift as their photo-$z$ cannot be calibrated. As a result, the clustering strength is diluted and some spurious cross-correlation signal appears resulting in a decreased FoM compared to a shallower sample with better photometric redshifts.

To illustrate this effect, in \Cref{fig:Nofz_3_bins_VIS_cuts} we show the redshift distribution of three tomographic bins for three samples with galaxies down to \VIS $<24.5$, $<25$, and with galaxies only between 24.5 and 25. Galaxies with \VIS between 24.5 and 25 are mostly outside their tomographic bin increasing the width of the distribution and diluting the signal.
We conclude that the \GCph probe is sensitive to the actual location of their tracer galaxies inside their tomographic bins. Both the photometric redshift performance and the number density are important contributing factors when performing cosmological inference with \GCph. When pushing to faint magnitudes, there is no improvement including galaxies that cannot be located in redshift. 

Let us now discuss the case where we add GGL to \GCph (right panel in \Cref{fig:FoM_vs_mag_pz}). We observe that increasing the sample size (moving along the $y$-axis) has a more significant impact on the improvements of the FoM than the photo-$z$ quality (changes along the $x$-axis). The greatest improvement, of about $50\%$ for the best photo-$z$ quality samples, takes place going from \VIS $<24$ to $24.5$. The second largest improvement is of about $25$ -- $30\%$ when adding objects from \VIS $<24.5$ to $25$. In the GGL case, source galaxies outside the tomographic bin of the lens galaxy contribute to the signal. The lensing kernel is quite extended in redshift and galaxies beyond the lens contribute to the signal with only a mild dependence on their precise redshift, making the photometric redshfit performance less important compared to the \GCph only case. On the other hand, the statistical nature of detecting the lensing signal makes the number density (and therefore the magnitude limit cut) a more important factor in determining the GGL cosmological inference power. 


In the FoM grid, we find a counter intuitive behaviour for some samples when combining \GCph and GGL (\Cref{fig:FoM_vs_mag_pz} right panel). If we compare the mid depth and mid depth \Euclid samples to the fiducial and optimistic samples at the same number density (along the $x$-axis), we find that the former pair gives better FoM constraints despite having larger photo-$z$ scatter. This is counter-intuitive as fewer galaxies are properly located in redshift and still the FoM cosmological constraints are slightly better. As we mentioned before, whenever the photo-$z$ performance degrades, more galaxies supposedly being in our tomographic bin belong to other bins. This effect can increase the effective number of sources for our lenses and thus boost the GGL signal. However, this is at the expense of reducing the cosmological constraining power of the \GCph probe. The interplay between these two effects is difficult to gauge. The GGL increase appears slightly more prominent when pushing to fainter magnitude limits that produce a sizeable increase in number density.   



The representativeness of the training sample also determines the photo-$z$ performance and thus the cosmological constraining power. For \GCph, if we check the difference in FoM between our fiducial sample, trained with a spectroscopic sample that has a completeness drop at faint \VIS, and the same photometric sample trained with a fully representative training sample (optimistic sample) we see a gain of about 1--2$\%$ in the FoM. Note that the spectroscopic incompleteness in this case is small and only affecting faint magnitudes, so the effect on the FoM is also small. This difference greatly increases when we compare the FoM performance of shallower samples and higher incompleteness in the spectroscopic training sample. If we compare the shallow depth sample that was trained with a sample that has a completeness drop in faint \VIS magnitude to the shallow depth inhomogeneous sample that was trained with a sample that is incomplete in the spectroscopic $n(z)$, the difference between FoMs can be up to $25\%$ for \GCph and $39\%$ for both probes combined.

Finally, we look at the difference due to the ground-based photometric depth. The difference between our fiducial and shallow depth cases may represent the change in depth to be achieved in the Southern and Northern hemispheres. For these cases the difference in cosmological constraint power is about $10\%$ at \VIS $<24.5$ for \GCph. This percentage reduces to $2\%$ if we also consider GGL.

\subsection{Impact on the cosmological parameters constraints}\label{subsec:Impact_cosmological_constraints}

\begin{figure*}
    \centering
	\includegraphics[width=\textwidth]{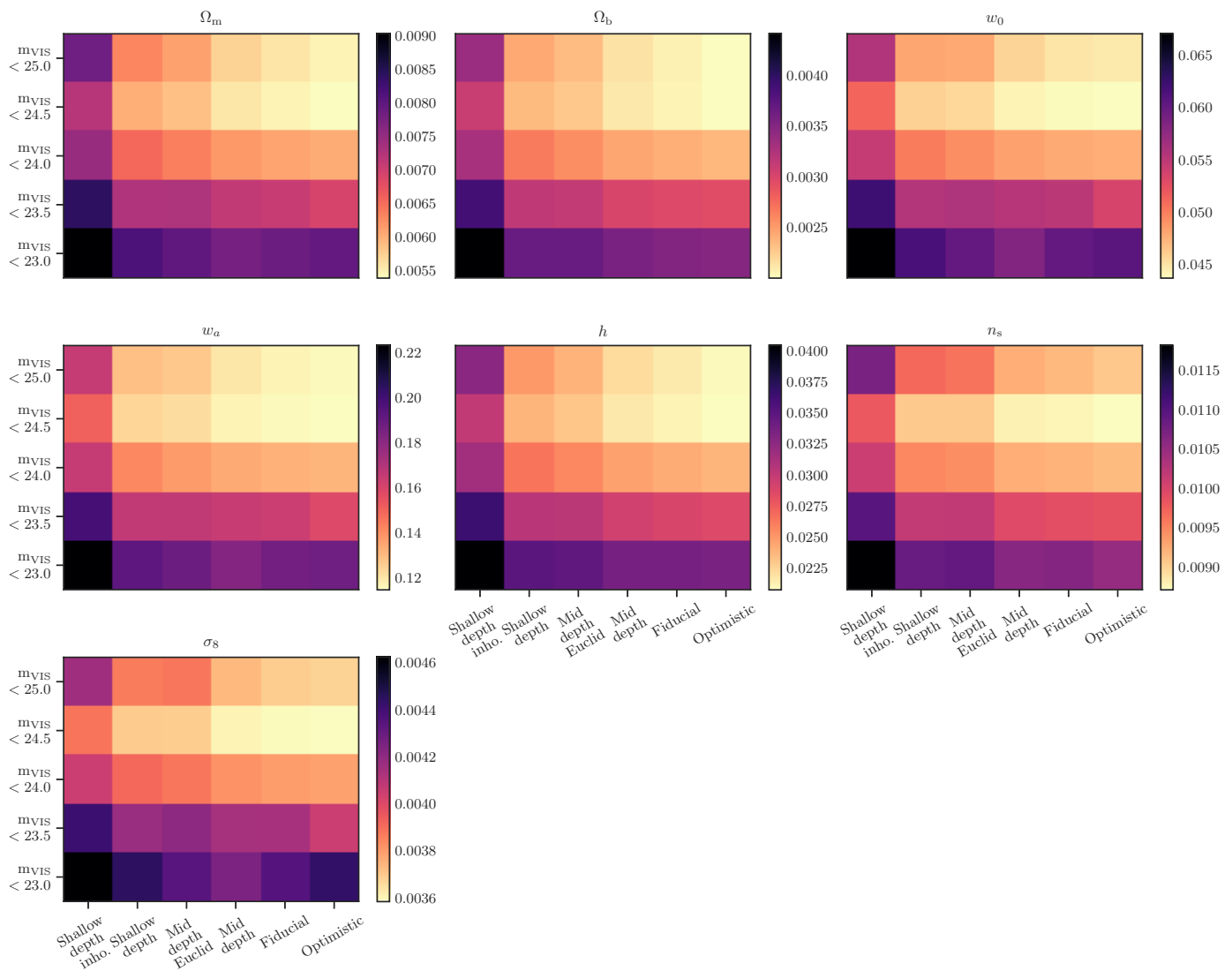}
    \caption{Uncertainties of the cosmological parameters for all the cases considered in \Cref{sec:Dep_photoz_sample_size} for \GCph.}
    \label{fig:Uncertainties_params_GCph}
\end{figure*}

\begin{figure*}
    \centering
	\includegraphics[width=\textwidth]{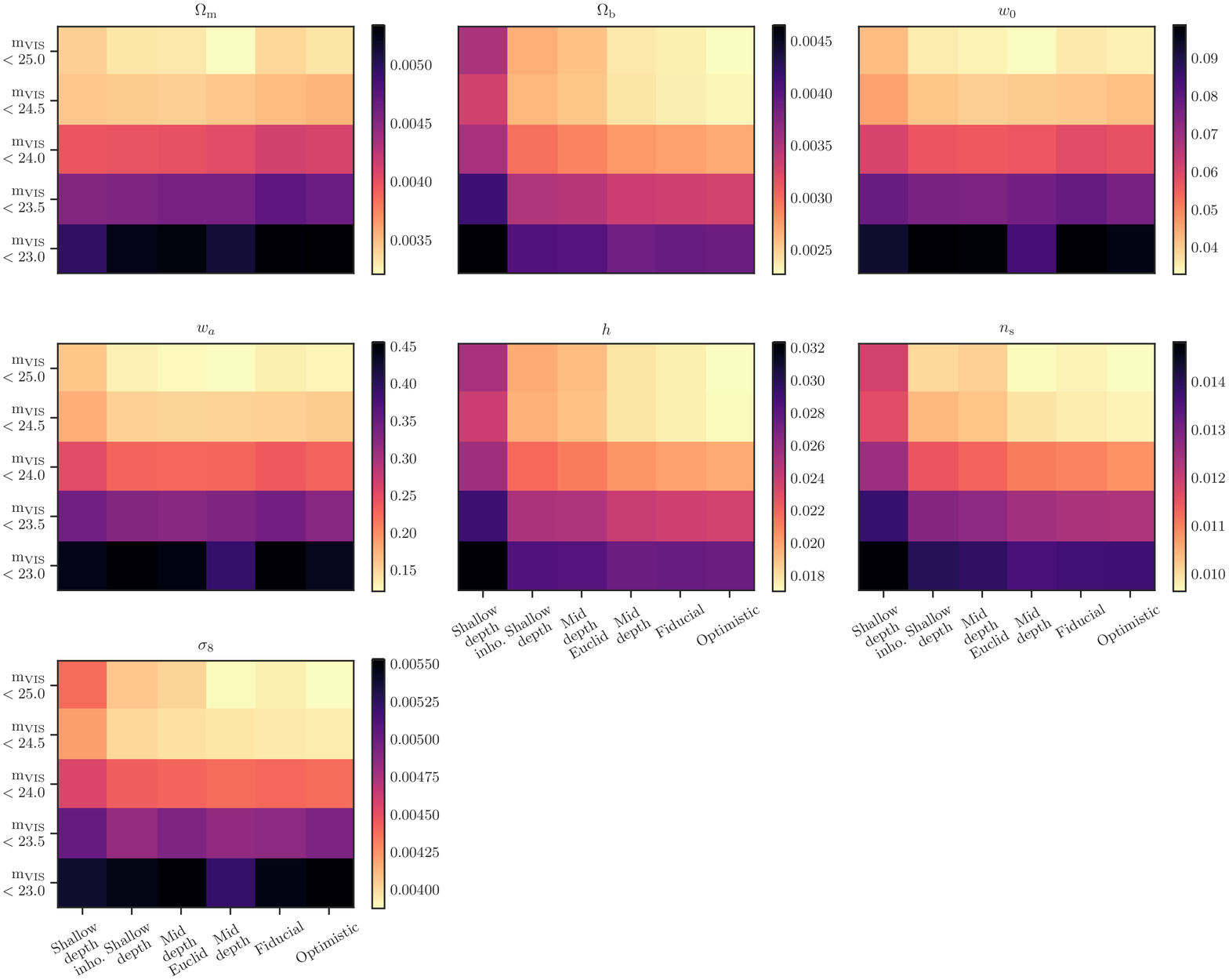}
    \caption{Uncertainties of the cosmological parameters for all the cases considered in \Cref{sec:Dep_photoz_sample_size} for combined \GCph and GGL.}
    \label{fig:Uncertainties_params_GCph_GGL}
\end{figure*}

We further investigate the forecasts of the constraints on the cosmological parameters by looking at the parameter uncertainties, $\sigma_{i} = ((\textbf{F}^{-1})_{ii})^\frac{1}{2}$, given by the square root of the diagonal elements of the inverse of the Fisher matrix. The uncertainties are computed for all the photometric samples defined in \Cref{sec:Samples} and for the different sample sizes. For visual clarity, we present the results in grid form in \Cref{fig:Uncertainties_params_GCph,fig:Uncertainties_params_GCph_GGL}.

In \Cref{fig:Uncertainties_params_GCph} we show the uncertainties for the \GCph probe. We can appreciate that, in general, the uncertainties have a similar behaviour to the FoM, where the sample down to \VIS $<24.5$ gives the higher FoM. However, there are parameters, such as $\Omega_{\rm b}$, $w_{\rm a}$, and $h$, that do not degrade as much their performance when going to the deeper \VIS $<25$ sample.

In \Cref{fig:Uncertainties_params_GCph_GGL} we show the uncertainties of the cosmological parameters when we combine the \GCph and GGL cosmological probes. Again, we see similar trends compared to the FoM case, but with minor changes in the behaviour of how the uncertainties in some of the parameters vary. 
The addition of galaxies, increasing the survey depth, and the improvement of the photo-$z$ performance produce lower uncertainties in the $\Omega_{\rm b}$ and $h$ parameters.
The reduction of the uncertainty obtained when considering the deepest \VIS $<25$ case compared to the \VIS $<24.5$ is minimal, though.

\begin{figure*}
    \centering
	\includegraphics[scale = 0.72]{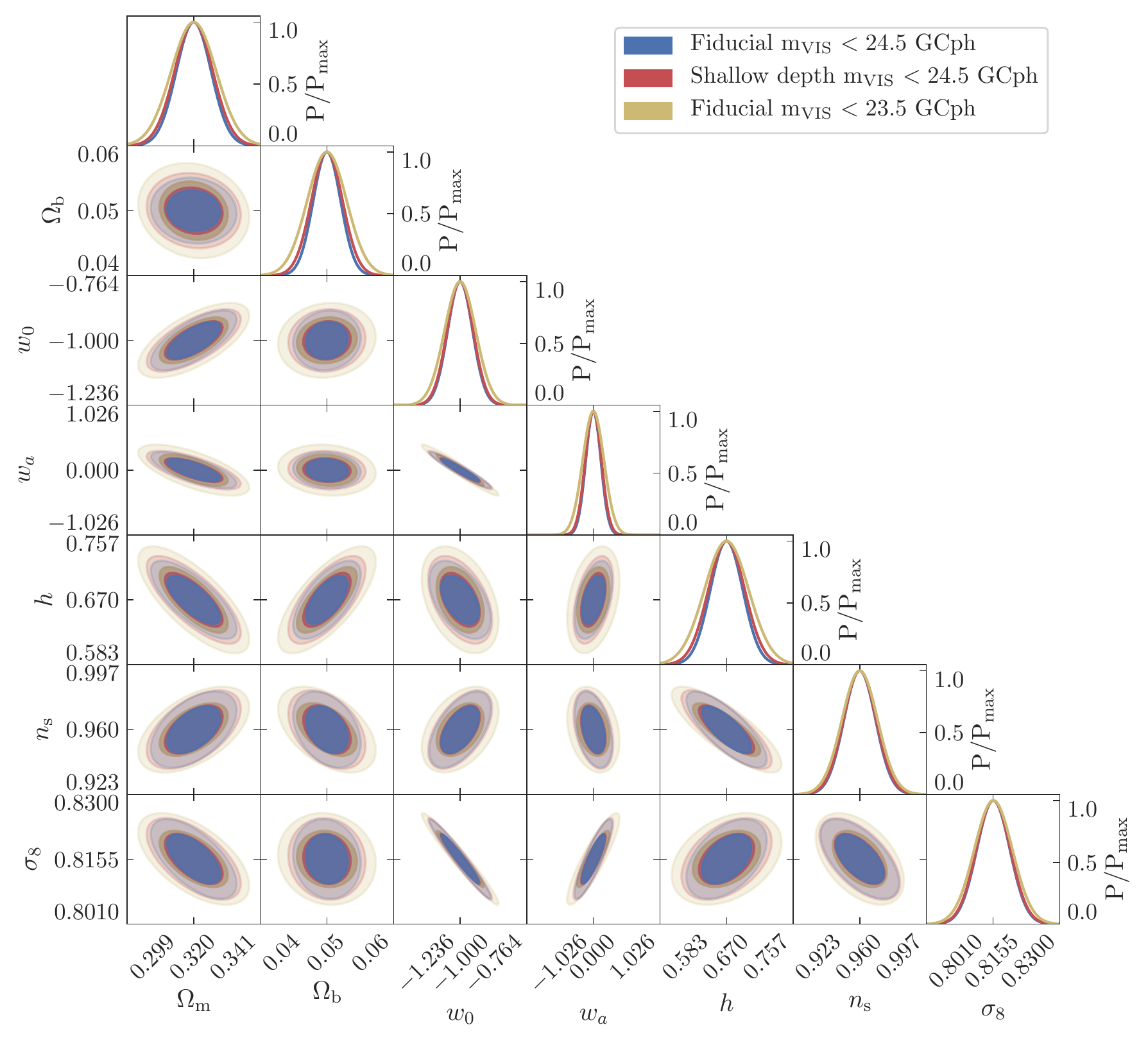}
	\includegraphics[scale = 0.72]{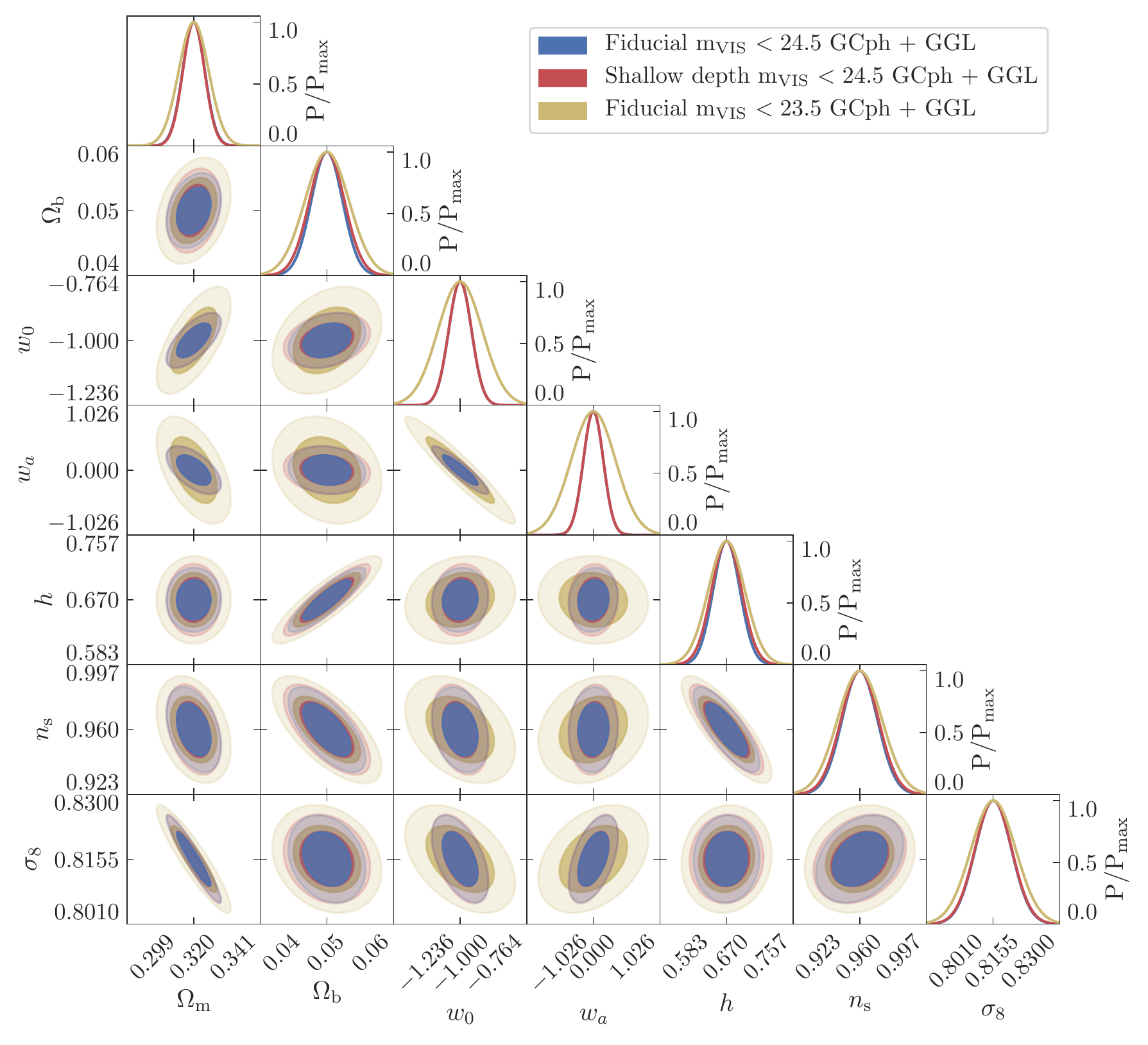}
    \caption{Fisher matrix contours for our fiducial sample down to \VIS $< 24.5$ (blue) and 23.5 (yellow), and the sample with ground-based photometry degraded by 1.75 magnitudes (red). \textit{Top panel}: For \GCph. \textit{Bottom panel}: For \GCph and GGL.}
    \label{fig:Contours_type_of_sample}
\end{figure*}

In addition to the values of the FoM and the uncertainties in the parameters, it is also informative to study the distribution of those uncertainties and the error contours in the determination of pairs of parameters. 
In \Cref{fig:Contours_type_of_sample} we present the confidence contour plots for our fiducial sample at \VIS $<24.5$ and $23.5$, to check how the number density affects the constraining power, and compare them to our shallow sample at \VIS $<24.5$, to see the impact of having a sample with shallower ground-based photometry. The contours for the \GCph case are shown in the upper panel and the \GCph and GGL case in the lower panel. For both probes we see that the fiducial sample gives the best constraints and the largest improvement is gained when the sample size increases. The increase in constraining power with sample size is more prominent in the \GCph and GGL combined case in general, and for the parameters that characterise dark energy, $w_{0}$ and $w_{a}$ in particular.

\subsection{Redshift distribution of the photometric redshift bins}
\label{sec:Closer_look_nz}

\begin{figure*}
    \centering
	\includegraphics[width=\textwidth]{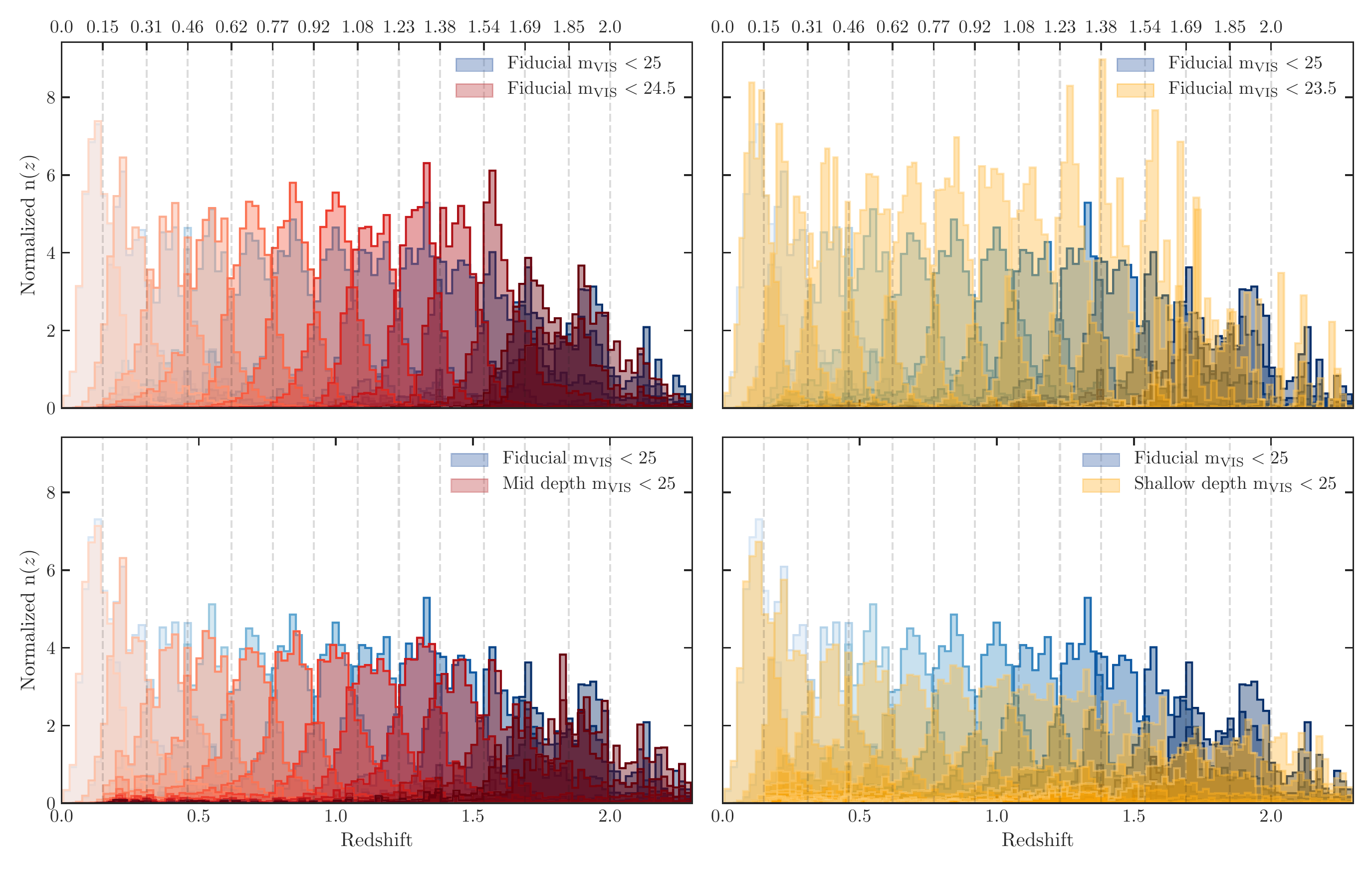}
    \caption{\textit{Top panels:} Redshift distribution ($z_{\text{mc}}$) of each tomographic bin for the fiducial sample at \VIS $< 25$ compared to the fiducial at \VIS $< 24.5$ (left) and $23.5$ (right).  \textit{Bottom panels:} Redshift distribution ($z_{\text{mc}}$) of each redshift bin for the fiducial sample compared to the mid depth (left) and shallow depth samples at \VIS $< 25$ (right).}
    \label{fig:Nofz_3_cases_comparison}
\end{figure*}

To better understand the behaviour of the FoM in \Cref{sec:Dep_photoz_sample_size} and the constraints in \Cref{subsec:Impact_cosmological_constraints}, we take a closer look at the $n(z)$ of some of the samples used to perform the study. 
In the top panel of \Cref{fig:Nofz_3_cases_comparison} we compare our fiducial photometric sample for \VIS cuts at $<25$, $<24.5$, and $<23.5$ to see the effects in the $n(z)$ when changing the magnitude limit and therefore the sample size. A shallower cut in magnitude removes objects at higher redshift. In the bottom panel of the figure we compare the $n(z)$ for the fiducial, mid depth, and shallow depth samples at \VIS $<25$ to see how the behaviour of the $n(z)$ changes with the depth of the ground-based photometry and therefore with the photo-$z$ performance.
Overall, the shallower the photometry, the larger the width of the $n(z)$ distributions, especially at higher redshift. 
This effect spuriously dilutes the correlation signal inside bins and increases the cross-correlation signal between bins, bringing down the \GCph constraining power. On the contrary, for the GGL case the widening of the redshift distributions is less important given the width of the lensing kernel. In addition, the effect of an increase in the number density dominates the performance of the FoM that in general increases with depth.  

In \Cref{tab:Number_sigma_z_bin_vis_cut_two_sample} we present quantitatively the number of objects per bin and the width of the $n(z)$ for the fiducial and shallow photometric samples, and for all the \VIS magnitude cuts.

\section{Summary and conclusions}\label{sec:conclusion}

Our primary goal is to study the cosmological constraints that can be derived from galaxy clustering studies of photometrically-selected samples using the combination of \Euclid and ground-based surveys. For that purpose we use the figure of merit, FoM, defined in \Cref{eq:FoM} as our performance metric.
We want to explore the impact of the ground-based photometry depth as well as the photometric redshift performance on the FoM constraints. To explore the photometric redshift performance, we vary both the survey depth and the spectroscopic sample available to train the photometric redshift algorithms. 
We use the Flagship simulation to create realizations of the expected observed magnitudes and their errors for the survey depths under study. 
To add a layer of realism to the study, we have computed the photo-$z$ using the machine learning code DNF in order to obtain a realistic photo-$z$ estimation for each of the photometric samples under study. We have also tried to mimic the training of the photo-$z$ method using spectroscopic samples with different completeness levels.  Given the scaled degradation of the photometric quality among the samples, we obtain a gradient of photo-$z$ quality. We choose as our fiducial sample the one corresponding to the photometric depth expected to be available in the Southern hemisphere with a survey like Rubin-LSST. We perform our FoM analysis using the same Fisher forecast formalism as in \citetalias{IST:paper1}.

First, we study the optimisation of the FoM with respect to the number and type of tomographic bins. We normalise our results to the case of ten bins with equal width since this is the specifications used in \citetalias{IST:paper1}. For this analysis we use the fiducial photometric sample defined in \Cref{sec:Samples}. \cref{fig:FoM_vs_nbins} shows the variations in the normalised FoM as a function of the number and type of bins. We find the best compromise for an optimal configuration to be:
\begin{itemize}
    \item Number of bins: A number slightly larger than ten is preferred. We adopt a default value of thirteen bins for our study. For bins with equal width, the FoM increases when moving from ten to thirteen bins by $35.4\%$ and $15.4\%$ for \GCph only and for \GCph + GGL, respectively. We find that a larger number of bins still provides an increase in the FoM for the \GCph only case. However, the photometric redshift scatter starts to be comparable to the bin width for such a large number of bins and our assumptions on how we train and compute the photo-$z$ may start to be too simplistic. 
    \item Type of bins: equal width. For the \GCph case the FoM increases by $30\%$ for thirteen bins with equal width compared to equipopulated bins. When combining with GGL the difference in the FoM as a function of bin type is almost negligible.
\end{itemize}
These results are in nice agreement with \citet{Kitching_2019} where they find similar conclusions of the optimum type of bins when optimizing the binning of photometric galaxy samples for cosmic shear analysis. The need of a larger number of bins, especially with good photometric redshift accuracy and the inclusion of intrinsic alignment parameters, to extract all the necessary information for cosmic shear is also found in \citet{Bridle_2007}. In this latter study, they also conclude that the model and freedom of the intrinsic alignment parameters greatly impact the FoM of dark energy.

We further study the dependence of the FoM on the quality of the photo-$z$ and the size of the sample. We study possible scenarios of complementary ground-based data for \Euclid that could be available in the Southern and Northern hemispheres and in the region in between. We take several magnitude limit cuts and generate realisations of the survey using the Flagship simulation. We also explore different possibilities of spectroscopy data available to train the photometric redshift techniques. We end up with a variety of samples with different number densities and photo-$z$ performance properties that try to encompass the  
possible samples that will be available for \Euclid analyses.
We compute the dark energy FoM for all these samples to study its variation.
Our results are summarised in \Cref{fig:FoM_vs_mag_pz} and \Cref{tab:FoM_vs_mag_pz_table}. For the \GCph case, we find a FoM of 713 for our fiducial sample with \VIS $<24.5$ (remember that galaxy bias is fixed for the \GCph case, providing larger absolute values for the FoM than in the combination of \GCph + GGL). The FoM improves with photo-$z$ quality and sample size. The trend with sample size or magnitude depth reverses when adding galaxies in a magnitude range (between $24.5$ and $25$ in our case) where photometric redshifts cannot be calibrated and are therefore of poor quality.  There is a faster increase of the FoM with sample size in those samples where the photo-$z$ performance is better. For example, in the optimistic, fiducial and mid depth cases increasing the sample size from \VIS 23.5 to 24 and from 24 to 24.5 leads to an increase in the FoM of about $20\%$.
When combining \GCph and GGL the FoM for the fiducial sample at \VIS $<24.5$ is 411. The FoM depends more strongly on the sample size (or survey depth) than on the photo-$z$ performance. The greatest FoM increase, of about $50\%$, takes place when adding galaxies from \VIS $<24$ to $24.5$.
The FoM has a weak dependence on the photo-$z$ performance. Generally, it improves with better photo-$z$ accuracy. 

The photometric redshift performance depends on the signal-to-noise of the photometry available and on the spectroscopic sample used in the photometric redshift algorithm. In our study, we use a machine learning technique, DNF.
The representativeness of the training sample has a significant influence on the photo-$z$ quality. The impact on the FoM is larger when the photometry is shallower.
For the optimistic photometry, the improvement in the FoM is minimal, 1--2$\%$, when we train the photometric redshifts with a representative subsample or with a subsample with a completeness drop at faint \VIS. This minimum variation is because the spectroscopic sample incompleteness in the second case only affects the very faintest galaxies. In the cases where the spectroscopy incompleteness is representative of a larger fraction of the galaxy sample the FoM variation is larger. For example, for our shallowest photometric sample, the relative variation in FoM when trained with an incomplete $n(z)$ and with just a completeness drop only at the faintest \VIS, can be of around $30\%$.

We also investigate the uncertainties in the constrains on our cosmological parameter set across the photo-$z$ quality and sample density space. Cosmological parameters present similar trends to those of the FoM. But there are small differences between the different parameters. For \GCph, in general the smallest uncertainty is achieved when we get the highest FoM, which is the optimistic sample at \VIS $<24.5$. However, $\Omega_{\rm b}$, $w_{\rm a}$ and $h$ get the smallest uncertainties for the same optimistic sample but for \VIS $<25$. The balance between the degradation of the photo-$z$ and the increase in number density affects these parameters slightly differently. For \GCph combined with GGL, the uncertainty in the cosmological parameters presents a similar behaviour to the FoM trends. The lowest uncertainty in the parameters is achieved when the number density is largest, at \VIS $<25$. The trend with photo-$z$ performance does not influence the level of uncertainty. In general the parameters are better constrained when the accuracy on the photo-$z$ determination is higher. However, for some parameters this trend is different in the deepest sample.

To conclude, there is significant gain in the FoM when using a larger number of redshift bins than the nominal ten bins choice of \Euclid, especially for \GCph. 
We study the effect that the accuracy of the photo-$z$s and the survey depth have on the FoM. When using the \GCph probe, the FoM increases with survey depth and with the reduction in photo-$z$ uncertainties. We study the influence of the training sample in the photo-$z$ performance and its implications on the FoM. We find than adding faint galaxies whose redshifts cannot be properly determined because there are no galaxies of those magnitudes in the training sample decreases the FoM.
For the combination of the \GCph and GGL probes, there is even more gain on the cosmological constraining power when using larger samples than for \GCph alone. 
The photo-$z$ quality has slightly less impact on the FoM than for \GCph alone.
In general for the combination of probes, the number density has a stronger influence on the FoM than the photo-$z$ accuracy.


\appendix 

\section{Additional table of $n(z)$ of photometric redshift bins}

In \Cref{sec:Closer_look_nz} we show the $n(z)$ for the fiducial, mid depth, and shallow samples at \VIS $<25$. In this appendix we present a detailed table containing the number of galaxies and the dispersion of $n(z)$ at each bin for the fiducial and shallow samples for all magnitude cuts. 


\begin{sidewaystable*}
	\centering
	\caption{Number of objects and dispersion, $\sigma$, for each redshift distribution within our fiducial redshift binning and for each \VIS cut. The table lists results which correspond to the fiducial sample and the shallow sample whose ground-based photometry is degraded by 1.75. The photo-$z$ of both samples are trained with a spectroscopic sample that has a completeness drop in \VIS.}
	\label{tab:Number_sigma_z_bin_vis_cut_two_sample}
    \begin{tabular}{l@{\hspace{0.25cm}}l@{\hspace{0.25cm}}lllllllllllll}
		\multicolumn{15}{c}{\textbf{Fiducial sample}}  \\ 
        \noalign{\smallskip}
        \hline
        \hline
        \noalign{\smallskip}
        \VIS & $z$ bin  &   0.0--0.15 &  0.15--0.31 &  0.31--0.46 &  0.46--0.62 &  0.62--0.77 &  0.77--0.92 &  0.92--1.08 &  1.08--1.23 &  1.23--1.38 &  1.38--1.54 & 1.54--1.69 & 1.69--1.85 &  1.85--2  \\
        \noalign{\smallskip}
        \hline
        \noalign{\smallskip}
\multirow{2}{*}{25} & Num. &  1110056 &   4172539 &   7690737 &   7742576 &   7525914 &   7295429 &   6241444 &   5981201 &   4223985 &   2982428 &   1921137 &   1449159 &   976552 \\
   & $\sigma$ &    0.073 &     0.103 &     0.107 &     0.122 &     0.141 &     0.162 &     0.187 &     0.183 &     0.203 &     0.223 &     0.249 &     0.233 &    0.217 \\
        \noalign{\smallskip}
        \cline{1-15}
        \noalign{\smallskip}
\multirow{2}{*}{24.5} & Num. &  1097615 &   3776899 &   5964220 &   5558628 &   5592743 &   5403185 &   4346079 &   4094858 &   2607993 &   1561814 &    755690 &    520928 &   363096 \\
   & $\sigma$ &    0.068 &     0.091 &     0.091 &     0.098 &     0.087 &     0.085 &      0.09 &     0.086 &     0.092 &     0.102 &     0.137 &     0.152 &    0.156 \\
        \noalign{\smallskip}
        \cline{1-15}
        \noalign{\smallskip}
\multirow{2}{*}{24} & Num. &  1043816 &   3109259 &   4465491 &   4274695 &   4344582 &   3720890 &   2404484 &   1773883 &    859017 &    355151 &    118249 &     69939 &    44695 \\
   & $\sigma$ &    0.061 &     0.078 &     0.078 &     0.079 &     0.069 &     0.068 &      0.07 &     0.069 &     0.073 &     0.082 &     0.117 &     0.139 &    0.145 \\
        \noalign{\smallskip}
        \cline{1-15}
        \noalign{\smallskip}
\multirow{2}{*}{23.5} & Num. &   932348 &   2439621 &   3418339 &   3449593 &   3274820 &   2360571 &   1241933 &    670591 &    207132 &     55555 &     12604 &      4697 &     2005 \\
   & $\sigma$ &    0.055 &      0.07 &     0.069 &     0.069 &     0.062 &     0.062 &     0.065 &     0.064 &     0.069 &     0.088 &     0.129 &     0.168 &    0.189 \\
        \noalign{\smallskip}
        \cline{1-15}
        \noalign{\smallskip}
\multirow{2}{*}{23} & Num. &   789691 &   1902902 &   2705811 &   2746874 &   2269715 &   1369104 &    555385 &    189037 &     33808 &      6408 &       887 &       253 &       75 \\
   & $\sigma$ &    0.051 &     0.065 &     0.064 &     0.063 &     0.059 &     0.058 &     0.063 &     0.064 &     0.077 &     0.105 &     0.165 &     0.221 &    0.151 \\
		\noalign{\smallskip}
		\hline
		\noalign{\smallskip}
        \multicolumn{15}{c}{\textbf{Shallow sample}} \\
        \noalign{\smallskip}
        \hline
        \noalign{\smallskip}
        \noalign{\smallskip}
\multirow{2}{*}{25} & Num. &   978190 &   3672690 &   7897148 &   8034871 &   7918102 &   7685291 &   6333228 &   5904729 &   4162337 &   3054888 &   1811004 &   1199176 &   766188 \\
   & $\sigma$ &    0.128 &     0.142 &      0.15 &      0.18 &     0.202 &     0.218 &     0.248 &     0.253 &     0.291 &     0.314 &     0.359 &     0.367 &    0.354 \\
        \noalign{\smallskip}
        \cline{1-15}
        \noalign{\smallskip}
\multirow{2}{*}{24.5} & Num. &   914714 &   3258214 &   6223068 &   5612186 &   5749345 &   5737669 &   4481509 &   4082193 &   2508584 &   1632392 &    763071 &    456446 &   311625 \\
   & $\sigma$ &    0.076 &     0.105 &     0.117 &     0.138 &     0.143 &     0.145 &     0.156 &     0.145 &     0.165 &     0.178 &     0.222 &     0.238 &    0.221 \\
        \noalign{\smallskip}
        \cline{1-15}
        \noalign{\smallskip}
\multirow{2}{*}{24} & Num. &   896794 &   2920536 &   4858119 &   4171760 &   4314051 &   3826551 &   2414157 &   1788392 &    821471 &    370273 &    114523 &     60248 &    36376 \\
   & $\sigma$ &    0.067 &     0.092 &     0.099 &     0.103 &     0.096 &     0.093 &     0.099 &     0.094 &     0.107 &     0.125 &      0.18 &     0.199 &    0.185 \\
        \noalign{\smallskip}
        \cline{1-15}
        \noalign{\smallskip}
\multirow{2}{*}{23.5} & Num. &   850214 &   2389463 &   3699364 &   3297370 &   3260529 &   2395127 &   1224782 &    684043 &    195546 &     55925 &     10649 &      3980 &     1659 \\
   & $\sigma$ &    0.062 &     0.083 &     0.086 &     0.083 &     0.076 &     0.075 &     0.079 &     0.076 &     0.087 &     0.101 &     0.162 &     0.193 &    0.198 \\
        \noalign{\smallskip}
        \cline{1-15}
        \noalign{\smallskip}
\multirow{2}{*}{23} & Num. &   760306 &   1879125 &   2851850 &   2643698 &   2273819 &   1380422 &    546639 &    195193 &     30615 &      5996 &       681 &       260 &       80 \\
   & $\sigma$ &    0.057 &     0.074 &     0.075 &     0.071 &     0.067 &     0.067 &     0.071 &     0.072 &     0.088 &     0.106 &     0.185 &     0.166 &    0.229 \\
   		\noalign{\smallskip}
		\hline
		\hline
	\end{tabular}
\end{sidewaystable*}

\begin{acknowledgements}
A.~Pocino acknowledges financial support from the Secretaria d'Universitats i Recerca del Departament d'Empresa i Coneixement de la Generalitat de Catalunya with additional funding from the European FEDER/ERF funds, L’FSE inverteix en el teu futur. I.~Tutusaus acknowledges support from the Spanish Ministry of Science, Innovation and Universities through grant ESP2017-89838, and the H2020 programme of the European Commission through grant 776247. S.~Camera acknowledges support from the `Departments of Excellence 2018-2022' Grant awarded by the Italian Ministry of Education, University and Research (\textsc{miur}) L.~232/2016. S.
~Camera is supported by \textsc{miur} through Rita Levi Montalcini project `\textsc{prometheus} -- Probing and Relating Observables with Multi-wavelength Experiments To Help Enlightening the Universe's Structure'. A.~Pourtsidou is a UK Research and Innovation Future Leaders Fellow, grant MR/S016066/1.
\AckEC
\end{acknowledgements}

\bibliographystyle{aa}
\bibliography{ref_paper.bib}

\end{document}